\pdfoutput=1
\documentclass[11pt,twoside,a4paper,cmspaper,final,collab]{cms-tdr}
\def\svnVersion{86af171}\def\svnDate{2024/08/16}\def\cmsCernNoTag{CERN-EP-2024-043}\def\cmsCernDate{\today}\def\cmsMessage{Published in the Journal of High Energy Physics as \href{http://dx.doi.org/10.1007/JHEP07(2024)293}{\doi{10.1007/JHEP07(2024)293}.}}
\begin{document}\cmsNoteHeader{HIG-21-005}

\providecommand{\cmsResize}[1]{\resizebox{\textwidth}{!}{#1}}
\newcommand{\HH}{\ensuremath{{\PH\PH}}\xspace}
\newcommand{\ppHHbbgg}{\ensuremath{\Pp\Pp\to\PH\PH\to\gamma\gamma\bbbar}\xspace}
\newcommand{\Zee}{\ensuremath{\PZ \to \Pe\Pe }\xspace}
\newcommand{\Zmumu}{\ensuremath{\PZ \to \PGm\PGm}\xspace}
\newcommand{\HHbbgg}{\ensuremath{\PH\PH\to\gamma\gamma\bbbar}\xspace}
\newcommand{\mHH}{\ensuremath{m_{\PH\PH}}\xspace}
\newcommand{\Hgg}{\ensuremath{\PH\to\gamma\gamma}\xspace}
\newcommand{\Hbb}{\ensuremath{\PH\to \bbbar}\xspace}
\newcommand{\HWW}{\ensuremath{\PH\to \PWp\PWm}\xspace}

\newcommand{\ggH}{\ensuremath{\Pg\Pg\PH}\xspace}
\newcommand{\VH}{\ensuremath{\PV\PH}\xspace}
\newcommand{\VBFH}{\ensuremath{\mathrm{VBF}\ \PH}\xspace}
\newcommand{\ttH}{\ensuremath{\ttbar\PH}\xspace}

\newcommand{\sigmaHH}{\ensuremath{\sigma_{\PH\PH}\xspace }}
\newcommand{\sigmaVBF}{\ensuremath{\sigma_{\text{VBF}~\PH\PH}\xspace}}
\providecommand*\sigmaGGF{\ensuremath{\sigma_{\Pg\Pg\text{F}}_{\PH\PH}}\xspace}
\providecommand*\GGF{\ensuremath{\Pg\Pg\text{F}}\xspace}
\providecommand*\VBF{\ensuremath{\text{VBF}}\xspace}
\newcommand{\BR}{\ensuremath{\mathcal{B}(\HH\to\bbgg)}}

\newcommand{\bbgg}{\ensuremath{\gamma\gamma\bbbar}\xspace}
\newcommand{\bb}{\ensuremath{\bbbar}\xspace}
\newcommand{\bbbb}{\ensuremath{\bbbar\bbbar}\xspace}
\newcommand{\bbtt}{\ensuremath{\tau\tau\bbbar}\xspace}
\newcommand{\bbZZ}{\ensuremath{\PZ\PZ\bbbar}\xspace}
\newcommand{\bbWW}{\ensuremath{\bbbar\PWp\PWm}\xspace}
\providecommand*\lambdahhh{\ensuremath{\lambda_{\PH\PH\PH}}\xspace} 
\providecommand*\klambda{\ensuremath{\kappa_\lambda}\xspace} 
\providecommand*\ktop{{\ensuremath{\kappa_\PQt}}\xspace} 
\providecommand*\CV{\ensuremath{\kappa_\text{V}}\xspace} 
\providecommand*\CVV{\ensuremath{\kappa_\text{2V}}\xspace} 
\providecommand*\cg{\ensuremath{\textit{c}_\text{g}}\xspace} 
\providecommand*\cgg{\ensuremath{\textit{c}_{2\text{g}}}\xspace} 
\providecommand*\ctwo{\ensuremath{\textit{c}_2}\xspace}

\newcommand{\METx}{\ensuremath{E_{x}^\text{miss}}\xspace}
\newcommand{\METy}{\ensuremath{E_{y}^text{miss}}\xspace}
\newcommand{\HTmiss}{\ensuremath{\HT^\text{miss}}\xspace}

\newcommand{\PHiggs}{\PH}
\newcommand{\Pbottom}{\PQb}
\newcommand{\APbottom}{\PAQb}
\newcommand{\Ptop}{\PQt}
\newcommand{\APtop}{\PAQt}
\newcommand{\Pquark}{\PQq}
\newcommand{\APquark}{\PAQq}
\newcommand{\Pgluon}{\Pg}
\newcommand{\SD}{\ensuremath{\mathrm{SD}}}
\newcommand{\WZ}{\ensuremath{\PW\PZ}}
\newcommand{\ZZ}{\ensuremath{\PZ\PZ}}
\newcommand{\WW}{\ensuremath{\PW\PW}}
\newcommand{\tW}{\ensuremath{\Ptop\PW}}
\newcommand{\tZ}{\ensuremath{\Ptop\PZ}}
\newcommand{\tV}{\ensuremath{\Ptop\PV}}
\newcommand{\ttW}{\ensuremath{\Ptop\APtop\PW}}
\newcommand{\ttWW}{\ensuremath{\Ptop\APtop\PW\PW}}
\newcommand{\ttZ}{\ensuremath{\Ptop\APtop\PZ}}
\newcommand{\ttV}{\ensuremath{\Ptop\APtop\PV}}
\newcommand{\qqH}{\ensuremath{\Pquark\Pquark\PHiggs}}
\newcommand{\WH}{\ensuremath{\PW\PHiggs}}
\newcommand{\ZH}{\ensuremath{\PZ\PHiggs}}
\newcommand{\tH}{\ensuremath{\Ptop\PHiggs}}
\newcommand{\mH}{\ensuremath{m_{\PHiggs}}}
\newcommand{\mtop}{\ensuremath{m_{\Ptop}}}
\newcommand{\fb} {\mbox{\ensuremath{\,\text{fb}}}\xspace}
\newcommand{\pb} {\mbox{\ensuremath{\,\text{pb}}}\xspace}
\newcommand{\ns} {\mbox{\ensuremath{\,\text{ns}}}\xspace}

\cmsNoteHeader{HIG-21-005}

\title{Search for Higgs boson pair production in the \texorpdfstring{\bbWW}{bbWW} decay mode in proton-proton collisions at \texorpdfstring{$\sqrt{s}=13\TeV$}{sqrt(s)=13 TeV}}

\date{\today}

\abstract{A search for Higgs boson pair (\HH) production with one Higgs boson decaying to two bottom quarks and the other to two \PW bosons are presented. The search is done using proton-proton collisions data at a centre-of-mass energy of 13\TeV, corresponding to an integrated luminosity of 138\fbinv recorded by the CMS detector at the LHC from 2016 to 2018. The final states considered include at least one leptonically decaying \PW boson. No evidence for the presence of a signal is
  observed and corresponding upper limits on the HH production cross section are derived. The limit on the inclusive cross section of the nonresonant \HH production, assuming that the distributions of kinematic observables are as expected in the standard model (SM), is observed (expected) to be 14 (18) times the value predicted by the SM, at 95\% confidence level. The limits on the cross section are also presented as functions of various Higgs boson coupling modifiers, and anomalous Higgs boson coupling scenarios. In addition, limits are set on the resonant \HH production via spin-0 and spin-2 resonances within the mass range 250--900\GeV.
}

\hypersetup{
  pdfauthor={CMS Collaboration},
  pdftitle={Search for Higgs boson pair production in the bbWW decay mode in proton-proton collisions at sqrt(s)=13 TeV},
  pdfsubject={CMS},
  pdfkeywords={CMS, Higgs, HH, BSM}
}

\maketitle
\section{Introduction}
\begin{figure}[htb!]
  \centering
  \includegraphics[width=0.3\textwidth]{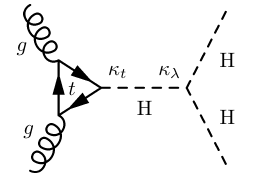}
  \includegraphics[width=0.3\textwidth]{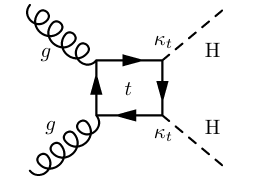}
  \caption{Leading-order Feynman diagrams of nonresonant Higgs boson pair production via gluon fusion in the standard model.}
  \label{fig:dihiggs-production-diagrams-ggf_SM}
\end{figure}

\begin{figure}[htb!]
  \centering
  \includegraphics[width=0.3\textwidth]{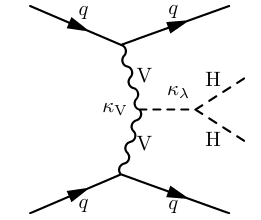}
  \includegraphics[width=0.3\textwidth]{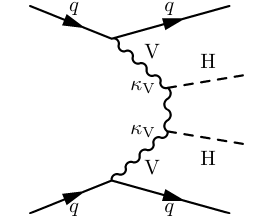}
  \includegraphics[width=0.3\textwidth]{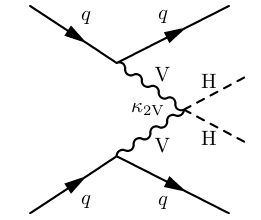}
  \caption{Leading-order Feynman diagrams of Higgs boson pair nonresonant production via vector boson fusion in the standard model.}
  \label{fig:dihiggs-production-diagrams-vbf}
\end{figure}

\begin{figure}[htb!]
  \centering
  \includegraphics[width=0.3\textwidth]{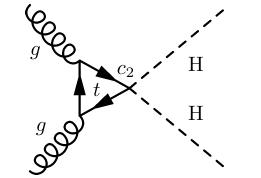}
  \includegraphics[width=0.3\textwidth]{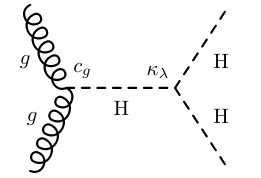}
  \includegraphics[width=0.3\textwidth]{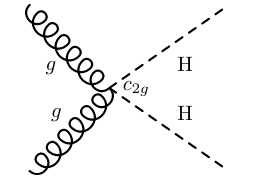}
  \caption{Leading-order Feynman diagrams of nonresonant Higgs boson pair production via gluon fusion with anomalous Higgs boson couplings.}
  \label{fig:dihiggs-production-diagrams-ggf_BSM}
\end{figure}

In 2012, the ATLAS and CMS Collaborations at the CERN LHC discovered a new particle with a mass of approximately 125\GeV~\cite{Higgs-Discovery_ATLAS, Higgs-Discovery_CMS, Higgs-Discovery_CMS_long}.
According to all current measurements, it is compatible with the standard model (SM) Higgs boson (\PHiggs)~\cite{Nature2022ATLAS, Nature2022, Sirunyan:2018koj, Aad:2019mbh}.
An important pending test of the electroweak symmetry breaking mechanism is the observation of Higgs boson pair (\HH) production.
At the LHC, pairs of SM Higgs bosons are primarily produced via gluon-gluon fusion (\GGF), with a cross section of 31.1$^{+2.1}_{-7.2}$\fb at 13\TeV centre-of-mass energy~\cite{2016_GGFXS,2019_GGFXS,Baglio:2020wgt,DiMicco:2019ngk,Grazzini:2018bsd}.
At leading order (LO), two destructively interfering Feynman diagrams contribute, the ``triangle diagram'' and the ``box diagram'', shown in Fig.~\ref{fig:dihiggs-production-diagrams-ggf_SM}.
The triangle-diagram gives direct access to the Higgs boson trilinear coupling \lambdahhh, which affects the shape of the Higgs field potential.
A secondary production mechanism for \HH events is the vector boson fusion (\VBF) shown in Fig.~\ref{fig:dihiggs-production-diagrams-vbf}.
While the cross section of the \VBF production is smaller, only 1.726$\pm$0.036\fb in the SM at 13\TeV~\cite{2018_VBFXS}, it gives experimental access to the quartic \mbox\PH\PH\PV\PV coupling (where \PV is a \PW or \PZ boson). The \HH production is also sensitive to other Higgs boson couplings, such as the \mbox\PH\PV\PV coupling. The Higgs boson couplings are described by their coupling modifiers, the ratio between the measured coupling strength and the prediction in the SM, noted with \ensuremath{\kappa}~\cite{LHCHiggsCrossSectionWorkingGroup:2013rie}. For example \klambda is the coupling modifier corresponding to the Higgs boson trilinear coupling and \ktop is the coupling modifier between a Higgs boson and a top quark. Beyond the SM, there may be additional diagrams contributing to \HH production that include couplings not predicted in the SM. The anomalous couplings studied in the present paper are denoted $c_i$ and shown in Fig.~\ref{fig:dihiggs-production-diagrams-ggf_BSM}. Here \ctwo corresponds to the coupling between two top quarks and two Higgs bosons, \cg corresponds to the coupling between a Higgs boson and two gluons, and \cgg corresponds to the coupling between two Higgs bosons and two gluons.

The \HH production could also be enhanced by resonant contributions through the production of a new heavy resonance (\PX) decaying to a pair of Higgs bosons. Examples of such new resonances include a radion~\cite{Cheung:2000rw}, a heavy $\textrm{CP}$-even scalar in two-Higgs-doublet models~\cite{TwoHDM} and a spin-2 graviton in the bulk Randall--Sundrum model~\cite{Randall:1999ee, Goldberger:1999uk}.

This paper describes a search for nonresonant and resonant \HH production in the decay channel to a pair of \Pbottom quarks and a pair of \PW bosons. The search is carried out by analysing proton-proton ($\Pp\Pp$) collision data recorded by the CMS detector~\cite{Chatrchyan:2008zzk} from 2016 to 2018 at 13\TeV centre-of-mass energy with single- and double-lepton triggers. The data sample corresponds to an integrated luminosity of 138\fbinv~\cite{LUM-17-003,LUM-17-004,LUM-18-002}.

The $\HH\to\bbWW$ decay has the second-largest branching fraction, following the $\HH\to\bbbb$ decay.
We consider events with at least one \PW boson decaying to an electron or a muon. Higgs boson decays to a pair of tau leptons with subsequent decay of both tau leptons to electrons or muons, as well as Higgs boson decays to a pair of \PZ bosons resulting in a final state with two electrons or muons, are also considered as signal.
From here on in this paper, the term leptons is used for electrons and muons unless explicitly stated otherwise. The main backgrounds contributing to this final state are the top quark pair production ($\ttbar{+}\text{jets}$), followed by the Drell--Yan (DY) or $\PW{+}\text{jets}$ processes
depending on whether the second \PW boson decays leptonically or hadronically, respectively. In the latter case, events with misidentified leptons represent a sizeable contribution. Other SM processes contribute to a lesser extent, for example single top quark and multiboson ($\PV\PV$ and $\PV\PV\PV$) productions.

The combination of \HH nonresonant searches by the CMS Collaboration~\cite{Nature2022} sets an upper limit on the inclusive \HH production cross section observed (expected) at 3.4 (2.5) times the theoretical cross section predicted by the SM, at 95\% confidence level (\CL). The coupling modifier for the trilinear Higgs boson self-coupling, \klambda, is excluded outside the range $-1.25$~and 6.85, at 95\% \CL, using the modified frequentist \CLs method~\cite{CLS1,CLS2}. The coupling modifier for the quartic interaction between two Higgs bosons and two \PW or \PZ bosons, \CVV, is excluded outside the range 0.67 and 1.38, at 95\% \CL, which corresponds to an exclusion of the \CVV$=0$ hypothesis by 6.6 standard deviations, when all other couplings are assumed to be SM-like. These results include several \HH decays but not \bbWW, namely \bbgg, \bbbb, \bbtt, \bbZZ and final states with leptons.
The ATLAS Collaboration published a combination of \HH results, including \bbgg, \bbbb and \bbtt decays, constraining the observed (expected) \HH production to 2.4 (2.9) times the SM value, at 95\% \CL~\cite{ATLAS:2022jtk}.
The ATLAS Collaboration has also published a result on the $\HH\to\bbWW$ channel in final states with two leptons~\cite{ATLASbbWW2024}. This search constrained the observed (expected) inclusive \HH production cross section to be lower than 9.7 (16.2) times the SM cross section, at 95\% \CL. The most recent result on $\HH\to\bbWW$ by the CMS Collaboration~\cite{HIG-17-006} used only the data collected in 2016, corresponding to an integrated luminosity of 35.9\fbinv and reported an observed (expected) exclusion limit at 79 (89) times the value predicted by the SM. The analysis presented in this paper improves this result by up to a factor 5. Besides the additional data from 2017--2018 used in this paper, the previous analysis considered only the fully leptonic \HWW decays while here the semileptonic decay is considered as well. The case of a highly Lorentz-boosted \Hbb decay is also considered for the first time in this decay channel. The sensitivity is further improved by the use of a better algorithm~\cite{Bols:2020bkb} for identifying jets originating from b quarks and a different machine learning strategy for signal extraction.

This paper is structured as follows: the apparatus and the simulated samples are described in Section~\ref{detector_reco} and Section~\ref{Simulated_samples}. Section~\ref{reco} summarises
the physics object reconstruction and identification. Event selection and analysis strategy are discussed in Section~\ref{Event_selection}~and Section~\ref{DNN_categories}. We then discuss the background estimation and the systematic uncertainties
in Section~\ref{Background_estimation}~and Section~\ref{systematics}. Finally, Section~\ref{final_results}~presents the results, and Section~\ref{summary} the summary. A HEPData record is provided for the results~\cite{hepdata}.

\section{The CMS detector}
\label{detector_reco}

The central feature of the CMS apparatus is a superconducting solenoid of 6\unit{m} internal diameter, providing a magnetic field of 3.8\unit{T}. Within the solenoid volume, there is a silicon pixel and strip tracker, a lead tungstate crystal electromagnetic calorimeter (ECAL), and a brass and scintillator hadron calorimeter (HCAL), each composed of a barrel and two endcap sections. Forward calorimeters extend the pseudorapidity ($\eta$) coverage provided by the barrel and endcap detectors. Muons are measured in gas-ionisation detectors embedded in the steel flux-return yoke outside the solenoid.

A more detailed description of the CMS detector, together with a definition of the coordinate system used and the relevant kinematic variables, can be found in Ref.~\cite{Chatrchyan:2008zzk}.

Events of interest are selected using a two-tiered trigger system.
The first level, composed of custom hardware processors, uses information from the calorimeters and muon detectors to select events at a rate of around 100\unit{kHz} within a time interval of less than 4\mus~\cite{CMS:2020cmk}.
The second level, known as the high-level trigger, consists of a farm of processors running a version of the full event reconstruction software optimised for fast processing, and reduces the event rate to around 1\unit{kHz} before data storage~\cite{Khachatryan_2017}.
\section{Simulated samples}\label{Simulated_samples}

The parton showering, hadronisation processes, and decays of $\Pgt$ leptons, including polarisation effects, are modelled using the generator \PYTHIA8.230~\cite{Sjostrand:2014zea} with the tune \textsc{CUETP8M1}~\cite{PYTHIA_CUETP8M1tune_CMS} for the 2016 data-taking period, and with the tune \textsc{CP5}~\cite{CP5tune_CMS} for the 2017--2018 data-taking periods.
The simulated samples produced by \PYTHIA with the \textsc{CUETP8M1} tune use the \textsc{NNPDF3.0} parton distribution functions (PDFs), whereas the samples produced with the \textsc{CP5} tune use the \textsc{NNPDF3.1} set~\cite{NNPDF1,NNPDF3,Ball:2017nwa}. Finally, the samples produced by \MGvATNLO $2.2.2$~\cite{MadGraph5_aMCatNLO,Alwall_2007,Frederix_2012} and \POWHEG v2 ~\cite{POWHEG1,POWHEG2,POWHEG3}, use the \textsc{NNPDF3.1} set.

The response of the CMS detector is modelled using the \GEANTfour toolkit \cite{geant4}. Additional $\Pp\Pp$ interactions in the same or nearby bunch crossings, referred to as pileup, are simulated using \PYTHIA and overlaid on the simulated events using event weights so that the distribution of the number of collisions matches the data.

\subsection{HH signal modelling}
\label{sec:HHsimulation}

The \HH signal samples for nonresonant \GGF production are generated using next-to-leading-order (NLO) matrix elements implemented in the \POWHEG program.
These samples are produced in four benchmark hypotheses with varying values of the \klambda modifier (\klambda = 0, \klambda = 2.45, \klambda = 5.0, and \klambda = 1 (SM)), while the others are kept to their SM expected values.
The dependence of the \GGF \HH cross section on \klambda and \ktop can be obtained from three terms corresponding to the diagrams involving \klambda, \ktop, and the interference~\cite{Heinrich_2019}. Therefore we can model any kinematic distribution of the \GGF production over a large range of \klambda and \ktop values using a weighted sum of three of the four generated samples. Each weighted sum of samples is then normalised to the corresponding next-to-NLO (NNLO) cross section~\cite{Grazzini:2018bsd}.

In order to study further modified values for the SM couplings as well as couplings not present in the SM we use an event-based reweighting method. The reweighting is based on a parameterisation of the differential cross section on the generator-level invariant mass of the \HH system and the angular distance between the two Higgs bosons in the azimuthal plane, which are sufficient to characterize the hard scattering that only has two degrees of freedom.
It allows to access any combination of coupling modifiers (\klambda, \ktop, \ctwo, \cg, \cgg), even for values that were not used in the sample generation.

The modelling of the \VBF process follows the same principle.
In this case, the samples are generated at LO with \MGvATNLO. Seven benchmark samples are generated with varying values of coupling modifiers \klambda, \CV and \CVV. The cross section depends on six terms that are combinations of these three coupling modifiers. Accordingly, six of the seven generated samples are used in the weighted combination.

The simulated samples for resonant \HH production are produced using LO matrix elements implemented in the \MGvATNLO. The resonances are assumed to be in the mass range 250--900\GeV, have narrow width compared to the experimental resolution, and have spin 0 or 2.

\subsection{Background simulation}
\label{sec:Bkgsimulation}
{Simulated samples for the $\ttbar{+}\text{jets}$, single top quark production and WW processes are generated by \POWHEG at NLO.
  The simulated transverse momentum (\pt) spectrum of the top quarks is harder than the one observed in data.
  Therefore, we compute a correction which is then applied to the $\ttbar{+}\text{jets}$ Monte Carlo samples.
  The top quark \pt is weighted by the ratio of the NNLO theoretical cross section over the cross section obtained from simulation~\cite{Czakon_2017}.
  Single Higgs boson and $\PZ\PZ$ backgrounds are simulated either by \POWHEG or \MGvATNLO at NLO depending on the production mechanism and the subsequent decay.
  The DY, $\PW{+}\text{jets}$,  $\ttbar\PW{/}\PZ$ and $\PW\PZ$ and all $\PV\PV\PV$ processes are generated at NLO with \MGvATNLO.
  The DY and $\PW{+}\text{jets}$ backgrounds are modelled using ``inclusive'' samples, covering the whole phase space, and complementary DY and $\PW{+}\text{jets}$ samples binned in the multiplicity of jets at generator level. The ``stitching'' of the different DY and $\PW{+}\text{jets}$ samples is documented in Ref.~\cite{Ehataht_2021rkh}. Additional DY samples produced using LO matrix elements implemented in \MGvATNLO are used when training the machine learning algorithms implemented in this analysis in order to reduce the statistical uncertainty. The DY, $\PW{+}\text{jets}$, and $\ttbar{+}\text{jets}$ samples are normalised to cross sections computed at NNLO accuracy~\cite{FEWZ,TTbarXsectionNNLO,Zcrosssection_2018}.
  The $\PW\PGg{+}\text{jets}$, $t\PGg{+}\text{jets}$  and other rare processes are simulated at LO with \MGvATNLO.}
\section{Physics object reconstruction and identification}
\label{reco}

\subsection{Particle-flow algorithm}

The particle-flow (PF) algorithm~\cite{PRF-14-001} aims to reconstruct and identify each particle (PF candidate) in an event, with an optimised combination of information from the various elements of the CMS detector. These reconstructed particles are the so-called PF candidates and are classified as electrons, muons, photons, and charged or neutral hadrons.
The energy of electrons is determined from a combination of the track momentum at the primary vertex, the corresponding ECAL cluster energy, and the energy sum of all bremsstrahlung photons attached to the track. The momentum of muons is obtained from the curvature of the corresponding track. The energy of charged hadrons is determined from a combination of their momentum measured in the tracker and the matching ECAL and HCAL energy deposits, corrected for the response function of the calorimeters to hadronic showers. Finally, the energy of neutral hadrons is obtained from the corresponding corrected ECAL and HCAL energies.
The PF candidates are the starting point for further object identification and are used to build more complex objects like jets, and missing transverse momentum. The primary vertex is taken to be the vertex corresponding to the hardest scattering in the event, evaluated using tracking information alone, as described in Section 9.4.1 of Ref.~\cite{CMS-TDR-15-02}.

\subsection{Small-radius jets}
\label{sec:ak4jets}

{\tolerance=1000 Small-radius jets are reconstructed from PF candidates, using the anti-\kt clustering algorithm~\cite{Cacciari:2008gp,Cacciari:2011ma} with a distance parameter of $R = 0.4$.
Charged particles not originating from the primary vertex are excluded from the jet clustering.
The energy of reconstructed jets is calibrated as a function of jet \pt and $\eta$~\cite{CMS-DP-2021-033,JES_JER_2017}.
Corrections based on the area and energy density of the jet are applied in order to compensate for effects from pileup.
The jets selected in this analysis are required to satisfy the conditions $\pt> 25$\GeV and $\abs{\eta} < 2.4$, as well as selection criteria to remove jets adversely affected by instrumentation or reconstruction failure. In order to reduce the number of jets originating from pileup among the jets with $\pt < 50$\GeV, a set of criteria is applied to the compatibility of the tracks associated with the jet with the primary vertex, the topology of the jet shape, and the track multiplicity \cite{PUJID}. All selected jets are required not to overlap with electrons or muons passing the medium selection defined in Section~\ref{signal_leptons} within $\Delta R = \sqrt{(\Delta \eta)^2 {+} (\Delta \phi)^2} < 0.4$. We refer to these jets as ``small-radius jets''.

A deep neural network (DNN) based algorithm, \textsc{DeepJet}~\cite{Bols:2020bkb}, is applied to identify small-radius jets originating from the hadronisation of $\Pbottom$ quarks (``\Pbottom tagging''). The medium working point that yields an efficiency of 75\% for identifying jets from $\Pbottom$ quarks (\Pbottom jets), with a 1\% (10\%) misidentification rate for jets from light-flavour (charm) quarks and gluons, is used throughout this analysis~\cite{CMS-DP-2018-058}. This DNN-based algorithm exploits observables related to the long lifetime of $\Pbottom$ hadrons and the high charged particle multiplicity and mass of $\Pbottom$ jets compared to light quark and gluon jets. Corrections are derived from data enriched in \Pbottom jets, in order to account for the difference in data and simulation efficiencies, as a function of the jet \pt, $\abs{\eta}$ and the algorithm output score. They are applied to simulated events to improve the agreement with the data in the whole range of algorithm output scores~\cite{CMS-DP-2018-058}.
The estimation of the \Pbottom jet energy can be biased by the presence of neutrinos in semileptonic decays within the jet and by the detector response. To correct for this, a DNN regression algorithm is trained on jet composition and shape information and applied to b-tagged jets, improving the energy resolution by 12--15\%~\cite{CMS:2019uxx}.

To improve the sensitivity to the \VBF production with its forward jets, an additional category of jets (referred to as VBF jets) is considered with similar selections except for $\pt>30\GeV$ and $\abs{\eta} \leq 4.7$. During the 2017 data-taking period, jets and unclustered PF candidates with $2.650 <\abs{\eta} <3.139$ and $\pt < 60\GeV$ are rejected to reduce the effect of the noise in the ECAL endcap at high $\abs{\eta}$.}

\subsection{Large-radius jets}
\label{sec:ak8jets}

Decays of high-\pt Higgs bosons into a pair of \Pbottom quarks result in final states with large Lorentz boost and as a result, the \Pbottom jets can be overlapping, forming one jet with large $R$ (``large-radius jet'') and substructure (i.e. the two overlapping jets are ``subjets'' of the large-radius jet). These jets are reconstructed using the anti-\kt algorithm with a distance parameter $R = 0.8$. Contributions from pileup are reduced by weighting the PF particles used as input to the reconstruction of large-radius jets with the ``pileup-per-particle identification'' algorithm~\cite{Bertolini:2014bba}.

The large-radius jets are required to not be within $\Delta R < 0.8$ from leptons passing the medium selection defined in Section~\ref{signal_leptons}, and to have $\pt > 200$\GeV and $\abs{\eta} < 2.4$.
The two subjets are required to have $\pt > 20$\GeV. At least one of the subjets is required to have $\pt> 30$\GeV and pass the medium \Pbottom~tagging working point of the \textsc{DeepCSV} algorithm~\cite{CMS-BTV-16-002}, with a \Pbottom~jet efficiency of 68\% and misidentification rate of 1\% for light-flavour and gluon jets. Similarly to the small-radius jets, a correction derived from data is applied to simulated events to account for the differences in selection efficiencies associated with the medium working point. The soft-drop mass ($m_{\SD}$) of the large-radius jet is reconstructed using the modified mass drop tagger (also known as the ``soft-drop'' (SD)) algorithm~\cite{Dasgupta:2013ihk,Butterworth:2008iy}, with an angular exponent $\beta = 0$, soft-cutoff threshold $z_{\text{cut}} < 0.1$, and characteristic radius $R_0 = 0.8$. This quantity is designed to remove soft and wide-angle radiation from the large-radius jet and is required to be within the range $30 < m_{\SD} < 210$\GeV. 
The N--subjettiness~\cite{Thaler:2010tr} $\tau_{\mathrm{N}}$ denotes a quantity that measures the alignment of the jet energy along the axes of candidate subjets and is interpreted as the compatibility of a jet to have N subjets. The ratio $\tau_{2}/\tau_{1}< 0.75$ quantifies the compatibility of the large-radius jet with the two-prong structure expected from the decay of a $\PW$, $\PZ$ or Higgs boson into two quarks. This ratio is used in this analysis in the discrimination of $\Hbb$ decays of a Lorentz-boosted Higgs boson from quark and gluon jets.

\subsection{Electrons and Muons}
\label{sec:electron_and_muonId}

The electron and muon selection is performed in two stages. The first stage is the identification and isolation of genuine electron and muon candidates. The second stage is the selection of leptons specifically for the different aspects of this analysis, namely signal selection and background estimation or rejection.

\subsubsection {Electron and muon identification}
\label{sec:basic_electronId}

The first step of the electron identification is performed by a multivariate analysis (MVA) algorithm~\cite{Khachatryan:2015hwa,ElectronID} based on a boosted decision tree~\cite{TMVA} which is trained to discriminate electrons against jets. In this analysis, we use the selection threshold with 90\% efficiency for prompt electrons originating from the primary vertex.
Electron candidates arising from photon conversions are suppressed by requiring that the track is missing no more than one hit in the innermost layers of the silicon tracker and is not matched to a reconstructed conversion vertex.

The first step of the muon identification consists of linking track segments reconstructed in the silicon tracking detector with those in the muon system~\cite{Chatrchyan:2012xi}.
Quality requirements are applied on the multiplicity of hits, the number of matched segments and the quality of the global muon track fit, quantified by its normalized $\chi^{2}$.
The muon candidates used in the analysis are required to pass the ``loose" PF muon identification criteria~\cite{MuonID}, which guarantee more than 99\% efficiency over the entire $\eta$ range. The probability of pions or kaons to be misidentified as ``loose'' muons is about 0.2\% and 0.5\% respectively.

\subsubsection{Electron and muon isolation}
\label{sec:electronAndMuonIsolation}

Electrons and muons in signal events are expected to be isolated. Lepton isolation is defined as scalar \pt sum of all charged particles, neutral hadrons, and photons reconstructed within a narrow cone centred on the lepton direction.
The size $R$ of the cone is inversely proportional to the \pt of the lepton, causing increased efficiency for passing the isolation criteria for leptons reconstructed in events with overlapping jets due to high Lorentz boost or high hadronic activity. $R$ varies from 0.05 for leptons with $\pt>200\GeV$ to 0.20 for leptons with $\pt<50\GeV$.
Only charged particles originating from the lepton production vertex are considered in the isolation sum. Residual contributions of pileup to the neutral component of the isolation of the lepton are taken into account using effective-area corrections:

\begin{equation}
  I^{\text{lep}} = \sum_{\text{charged}} \pt + \max \left( 0, \sum_{\text{neutral}} \pt - \rho \, \mathcal{A} \, \left(\frac{R}{0.3}\right)^{2} \right),
  \label{eq:lepMiniIsolation}
\end{equation}

where $\rho$ represents the energy density of neutral particles reconstructed within the geometric acceptance of the tracking detectors, computed as described in Refs.~\cite{Cacciari:2008gn, Cacciari:2007fd}. The leptons considered are required to have $I^{\text{lep}}/\pt^{\text{lep}} < 0.4$. The effective area $\mathcal{A}$ is obtained from simulation by studying the correlation between $I^{\text{lep}}$ and $\rho$. It is determined separately for electrons and muons in bins of $\eta$.

\subsubsection{Signal lepton selection}
\label{signal_leptons}
The analysis utilises three different levels of lepton selection criteria with increasing tightness for electrons and muons, to which we refer to as the loose, medium and tight lepton selections.

The loose selection is used to remove lepton pair resonances. Requirements are the following: $\pt>5\GeV$ ($7\GeV$) and $\abs{\eta} < 2.5$ ($2.4$) for electrons (muons); isolation $I^{\text{lep}}/\pt^{\text{lep}}<0.4$; impact parameters of the lepton track with respect to the primary vertex, transverse $\abs{d_{xy}}<0.05$\cm and longitudinal $\abs{d_{z} }<0.1$\cm; significance of the impact parameter $d/\sigma_{d}<8$ in three dimensions (3D).

The medium lepton selection is used for removing the overlap between different types of objects, for certain variables in the event preselection and the misidentified-lepton background estimate based on control samples in data. The \pt of medium leptons is set to the $\pt^{\text{cone}}$, as this has been found to describe better the \pt of misidentified leptons. The $\pt^{\text{cone}}$~is defined as $0.9$ times the \pt of the nearest jet if they are within $\Delta R < 0.4$, otherwise as
$0.9 (\pt^{\text{lep}} {+} I^{\text{lep}})$, where $I^{\text{lep}}$ is the lepton isolation given by Eq.~(\ref{eq:lepMiniIsolation}). The $\pt^{\text{cone}}$~in general exceeds the \pt of the lepton as determined by the electron and muon reconstruction algorithms. Medium leptons are required to have $\pt^{\text{lep},\text{cone}}>10\GeV$ and the jet nearest to the lepton should fail the medium \Pbottom tagging working point. Medium electrons are further required to satisfy criteria similar to the ones applied at the trigger level. The requirements are: the width of the electron cluster in $\eta$-direction $\sigma_{i\eta i\eta}<0.011 (0.030)$ when $\abs{\eta}\leq 1.479$ ($\abs{\eta}> 1.479$); the ratio of energy associated to the electron in the HCAL to the energy in the ECAL $H/E<0.10$; the difference between the reciprocal of the electron cluster energy and the reciprocal of its track momentum $(1/E - 1/p)>-0.04$;
medium electron tracks are not allowed to miss any hits in inner pixel detector and to not originate from a photon conversion. Finally, only medium electrons not overlapping with medium muons within $\Delta R < 0.4$ are considered.

The tight lepton selection is used to select events in the signal region (SR). For this purpose, a lepton identification algorithm based on boosted decision trees is used, discriminating the prompt leptons from nonprompt and misidentified ones. Thereafter this will be referred to as the prompt-lepton MVA. The MVA is trained separately for electrons and muons~\cite{ttHmultilepton}. Several observables related to the lepton are used as input variables such as the $\pt^{\text{lep}}$, $\eta^{\text{lep}}$, relative
isolation $I^{\text{lep}}/\pt^{\text{lep}}$, $\abs{d_{xy}}$, and $\abs{d_{z} }$. The jet reconstruction and \Pbottom tagging algorithms are applied to the charged and neutral particles reconstructed in a cone around the lepton direction. The ratio of the lepton \pt to the reconstructed jet \pt and the component of the lepton momentum in a direction perpendicular to the jet direction are also used as inputs to the MVA. Tight leptons are required to have prompt-lepton MVA score greater than 0.5 (0.3) for muons (electrons). Muons are additionally required to pass the medium PF muon identification criteria as described in Ref.~\cite{MuonID}. Contrary to medium leptons, the \pt of tight leptons is not replaced by the $\pt^{\text{cone}}$ as these leptons are likely to be prompt.

\subsection{Missing transverse momentum}
\label{MET}

The missing transverse momentum vector \ptvecmiss is computed as the negative vector sum of the transverse momenta of all the PF candidates in an event and its magnitude is denoted as \ptmiss~\cite{CMS:2019ctu}. The \ptmiss is modified to account for corrections to the energy scale of the reconstructed jets in the event.

The variable \HTmiss is defined in the same way as \ptmiss, but considering only jets that fulfill the criteria described in Sections \ref{sec:ak4jets} and \ref{sec:ak8jets}, as well as electrons and muons passing the medium selection criteria, when evaluating the \pt sum. The observable \HTmiss has the advantage of being less sensitive to pileup, as soft hadrons that predominantly originate from pileup do not enter its computation, and can therefore be used as a complement to \ptmiss in a MVA algorithm.

This analysis also uses a linear combination, also referred to as a linear discriminant (LD), of \ptmiss and \HTmiss defined as $p_{\mathrm{T,LD}}^{\text{miss}} =0.6  \ptmiss {+} 0.4  \HTmiss$~\cite{ttHmultilepton}. The two observables are less correlated for events in which the \ptmiss arises from artificial effects compared to events with genuine \ptmiss. The jet energy corrections are propagated to \HTmiss and $p_{\mathrm{T,LD}}^{\text{miss}}$.

\section{Event selection}
\label{Event_selection}

The signature of the \bbWW signal is characterised by the decay of each of the Higgs bosons, \Hbb and \HWW. The \Hbb decay can be identified by the two jets originating from the hadronisation of each of the \Pbottom quarks. In order to identify \Hbb candidates, events are required to have at least one \Pbottom-tagged small-radius jet, accounting this way for inefficiencies in \Pbottom tagging. In case the Higgs boson is highly Lorentz-boosted, the two \Pbottom jets from the \Hbb decay are merged and they are reconstructed one single large-radius jet. Events are selected when they have one large-radius jet with least one of the subjets \Pbottom-tagged.

The VBF jet candidates must not overlap with the \Hbb candidates, either the small-radius jets within $\DR<0.8$, or the large-radius jet within $\DR<1.2$. Additionally, only pairs of jets with invariant mass $m_{jj}>500\GeV$ and separation $\Delta\eta_{jj}>3$ are considered. The leading pair in invariant mass determines the two VBF jets considered.

The two channels of this analysis are characterised by the decay products of the \HWW decay. The ``single-lepton'' channel targets events where only one \PW boson decays leptonically, while the  ``dilepton channel'' targets events where both \PW bosons do.

The first two steps of the event selection are the event preselection and trigger selection. The selected events are then evaluated using DNNs and categorised as described in Section \ref{DNN_categories}.

\subsection{Event preselection}
\label{sec:eventSelection_cleaning}
Events containing a pair of leptons passing the loose selection and with an invariant mass less than $12$\GeV are rejected as they are likely to originate from quarkonia decays. Events with a pair of loose-selection leptons, with opposite electric charge but same flavour, within 10\GeV of the mass of the \PZ boson (of 91.2\GeV~\cite{PDG}), are vetoed to remove DY and \ttZ~ backgrounds. To suppress effects related to beam halo, detector noise, \etc, the primary vertex in all events is required to have longitudinal distance from the collision point  $\abs{z_{\mathrm{vtx}}}$ less than 24\unit{cm}, radial distance $\abs{r}$ less than 2\unit{cm}, and at least four associated tracks.

\subsection{Trigger selection}
\label{sec:eventSelection_triggers}

The events selected in the single-lepton channel are required to pass either the single-electron or the single-muon trigger, based on the offline-reconstructed lepton flavour, selected as described in Section~\ref{sec:electron_and_muonId}. The \pt requirements applied in this analysis (Section~\ref{sec:eventSelection_single_lepton}) are chosen to be close to the trigger threshold (22--35\GeV) to reduce turn-on effects in the trigger selection efficiency. Residual turn-on effects are corrected for the simulated events using scale factors and corresponding systematic uncertainties.
In the dilepton channel, the acceptance for the \HH signal is increased by using a combination of single-lepton and dilepton triggers.
The dilepton triggers have a lower \pt threshold for the leading-\pt lepton (17--23\GeV) compared to the single-lepton triggers, which allows lowering the \pt thresholds for the electrons and muons reconstructed offline.

\subsection{Single-lepton channel}
\label{sec:eventSelection_single_lepton}

Events in the single-lepton channel are required to contain a lepton satisfying tight selection criteria with $\pt > 32$\GeV for an electron and $25\GeV$ for a muon.
Events containing a second lepton passing the tight selection criteria are vetoed to avoid overlap with the dilepton channel.

Events selected in the single-lepton channel are required to contain either at least three small-radius jets (Section~\ref{sec:ak4jets}) or at least one large-radius jet (Section~\ref{sec:ak8jets}) and at least one small-radius jet, which is separated from the large-radius jet by $\Delta R > 1.2$. At least one of the three small-radius jets, or in the latter case the large-radius jet, is required to be \Pbottom-tagged.
Overlap with the events selected by the $\HH\to\bbtt$ search is removed by vetoing events containing at least one hadronically decaying tau lepton, identified by the \textsc{DeepTau} algorithm~\cite{CMS:2022prd} as described in Ref.~\cite{CMS:2022hgz}.

\subsection{Dilepton channel}
\label{sec:eventSelection_dilepton}

Events selected in the dilepton channel must contain two leptons that pass the tight selection criteria and have opposite electric charges.
The leading lepton is required to have $\pt > 25$\GeV and the subleading one $\pt > 15$\GeV.
Events containing a third lepton passing the tight selection criteria are vetoed, to avoid overlap with the search for $\HH\to\bbZZ$ with four charged leptons in the final state.
The events are further required to contain at least either one \Pbottom-tagged large-radius jet or one \Pbottom-tagged small-radius jet.
\section{Analysis strategy}
\label{DNN_categories}

Events passing the single-lepton and dilepton selections are separated in different categories based on the signal purity, utilizing the \Hbb signal topology and fully connected DNN multiclassifiers, as described later in this section. The event categorisation is summarised in Tables~\ref{tab:Strategy_scheme_SL}~and~\ref{tab:Strategy_scheme_DL}.
To extract the \HH signal we perform a maximum likelihood fit in the asymptotic approximation~\cite{CMS-NOTE-2011-005} on the distribution of the DNN score, using the modified frequentist \CLs method~\cite{CLS1,CLS2}, simultaneously on the signal and background event categories.
To account for the effects of the systematic uncertainties, we include them as nuisance parameters (Section~\ref{systematics}) in the maximum likelihood fit.

Four DNNs are trained separately for the single-lepton and dilepton channels, and for the resonant and nonresonant signals. The DNNs for the nonresonant signal are trained on all available simulated signal events representing different coupling scenarios. For the resonant signal the DNNs are trained on all signal samples and the networks are parametrised~\cite{Baldi:2016fzo} according to the mass of the resonance decaying to the two Higgs bosons, which is provided in the training inputs for signal events and random values for background events in the same proportions. The background events are drawn from the simulated samples described in Section~\ref{sec:Bkgsimulation}.

The DNN architecture is complemented by a Lorentz Boost Network~\cite{LBN} acting as input preprocessor. This network takes as input the four-vectors of the reconstructed particles and creates additional observables (such as a two-particle invariant mass or angle difference in their centre-of-mass), which are then given as input to the DNN together with other variables related to the \HH signal topology. The combined Lorentz Boost Network and following dense networks are trained all together.

\begin{figure}[htb!]
  \centering
  \includegraphics[width=0.49\textwidth]{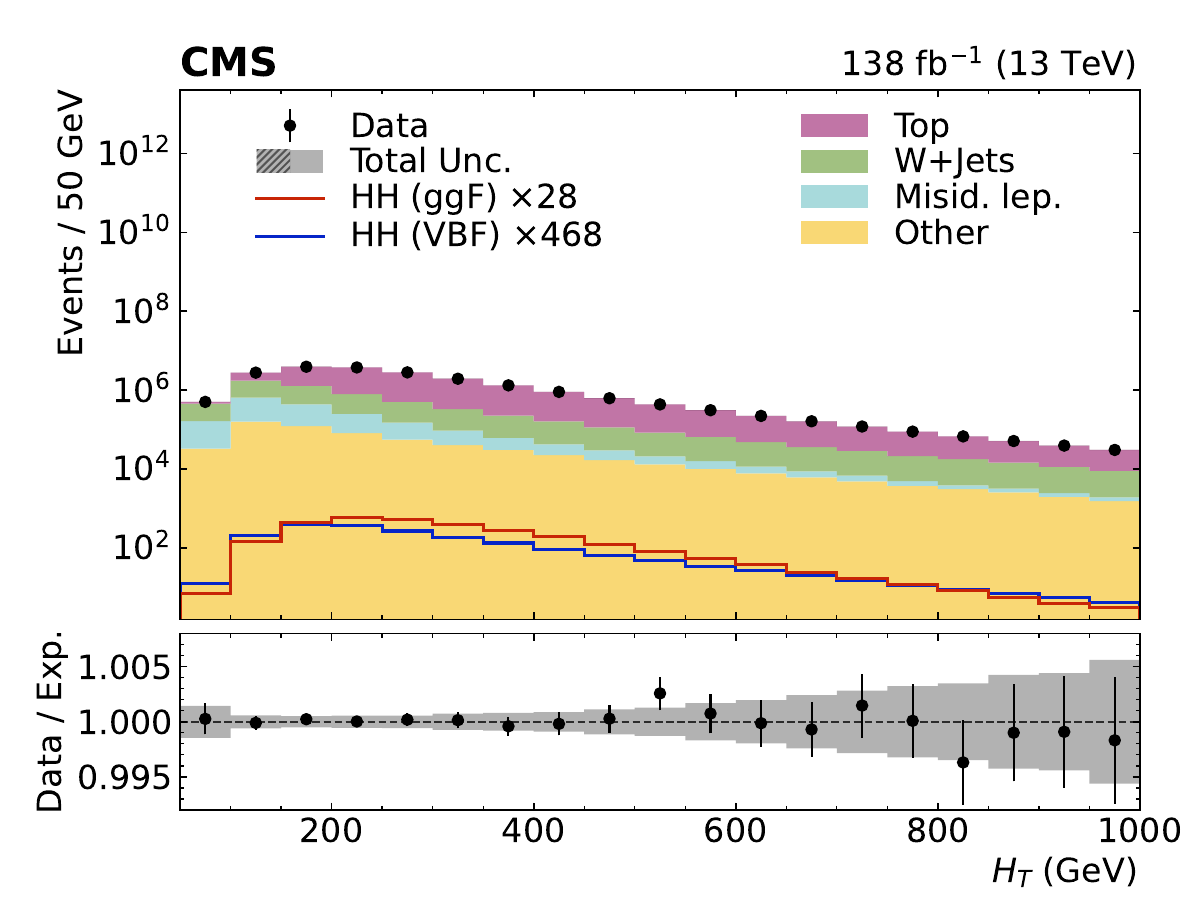}
  \includegraphics[width=0.49\textwidth]{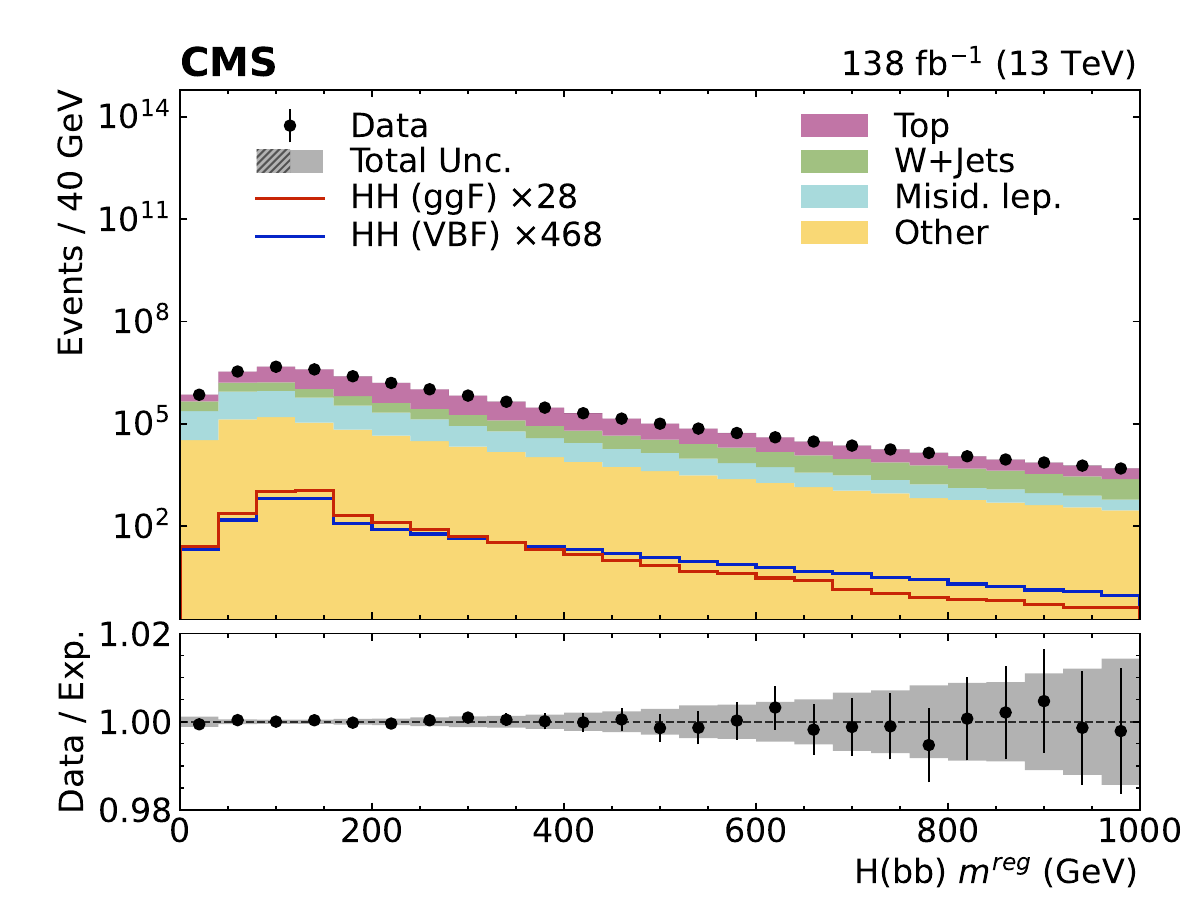}\\
  \includegraphics[width=0.49\textwidth]{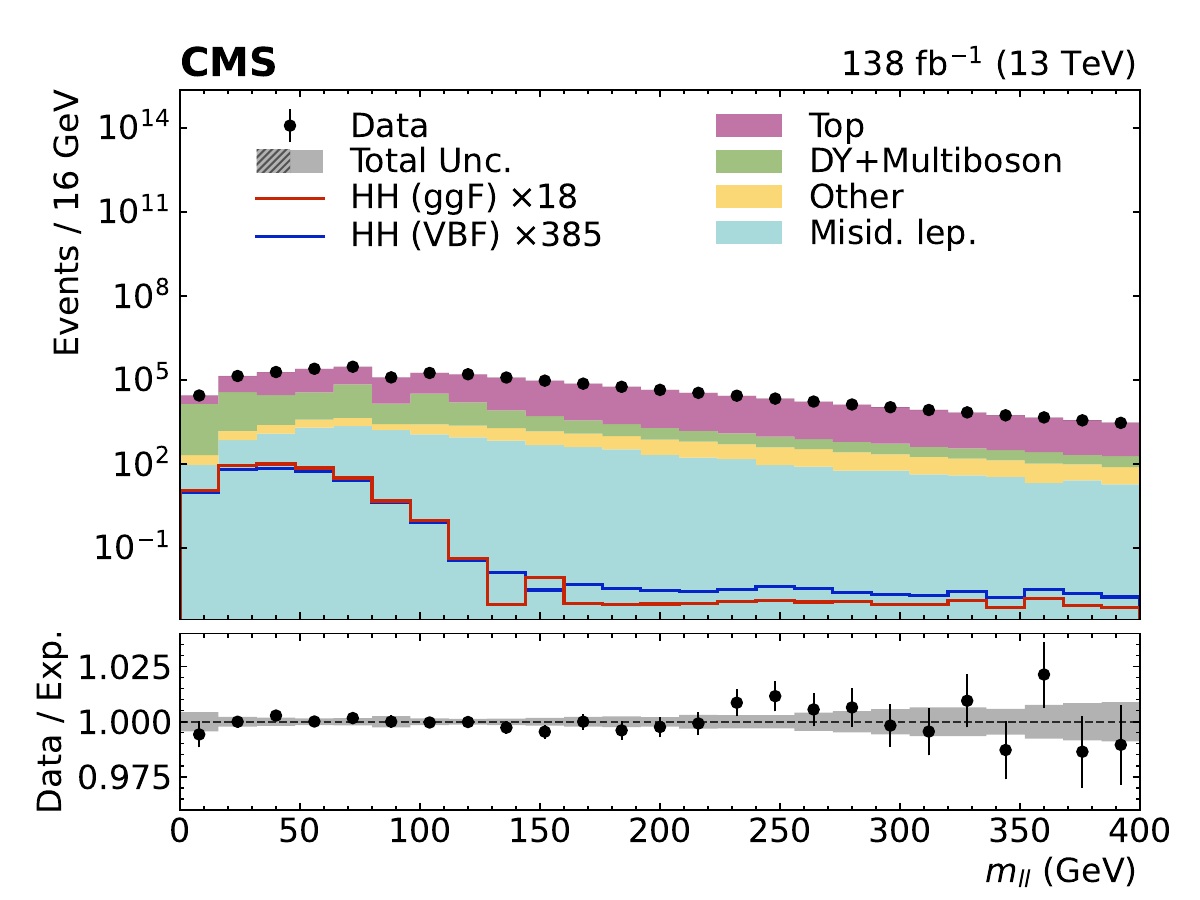}
  \includegraphics[width=0.49\textwidth]{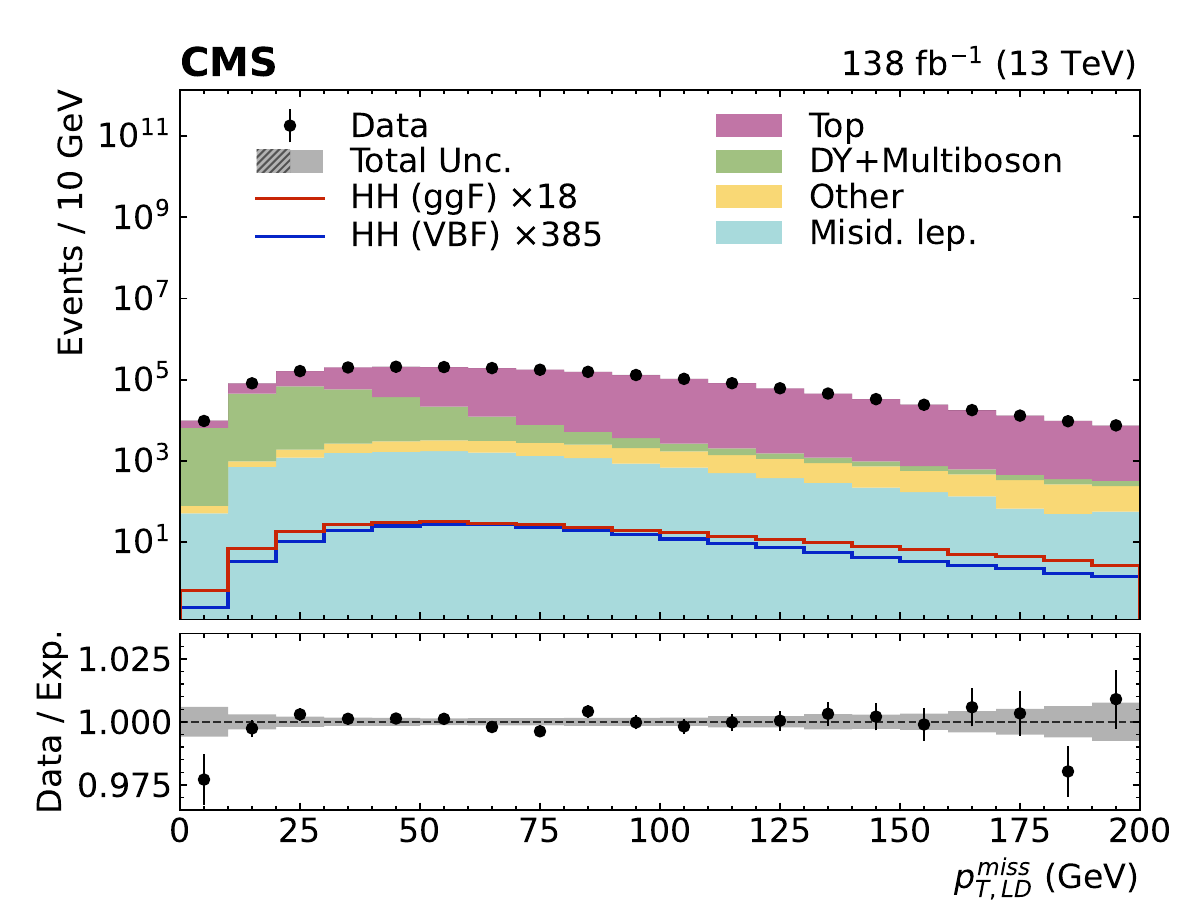}
  \caption{The distributions of some of the discriminants included in the DNN training for the single-lepton channel (upper) and the dilepton channel (lower). The distributions are shown after performing a maximum likelihood fit on the data for the variable pictured, using the same set of nuisance parameters (Section~\ref{systematics}) as in the likelihood fit used to extract signal. The variables are from upper left to lower right: the $H_{\mathrm{T}}$ variable, defined as the scalar sum of
    all selected jets \pt; the invariant mass of the two \Pbottom-tagged jets; the invariant mass of the two leptons; the $p_{\mathrm{T,LD}}^{\text{miss}}$, as defined in Section~~\ref{MET}. The signal shown is scaled to the observed upper limit on its cross section.}
  \label{fig:DNN_input}
\end{figure}

Two of the variables used as input to the DNN for each channel are shown in Fig.~\ref{fig:DNN_input}. The distributions are shown after performing a maximum likelihood fit on the data in the distribution of the variable displayed, using the same set of nuisance parameters (Section~\ref{systematics}) as in the likelihood fit used to extract the signal. The variables are from upper left to lower right: the \HT variable, defined as the scalar sum of all selected small-radius jets \pt; the invariant mass of the two \Pbottom-tagged jets after the regression correction (Section~\ref{sec:ak4jets}); the invariant mass of the two leptons; the $p_{\mathrm{T,LD}}^{\text{miss}}$ as defined in
Section~\ref{MET}. The highest ranking variable in terms of discrimination power is the \HT for both single-lepton and dilepton channels. The invariant mass of the two \Pbottom-tagged jets and the invariant mass of the two leptons are the second highest ranking variables for the two channels, respectively. The $p_{\mathrm{T,LD}}^{\text{miss}}$ ranks fifth for the dilepton channel, after the information about the flavour of the two leptons.

The DNNs are trained as multiclassifiers, which means that they learn each physics process (\ttbar, DY, etc.) separately, which in the machine learning context are referred to as ``classes". The multiclassifier assigns a score between $[0,1]$ for each class to each event. This score is related to the probability of an event belonging to the corresponding class. The event categorisation into background and signal categories is based on the DNN output scores; each event is assigned to the class with the highest probability. In the SM nonresonant case there are two signal categories as defined by the DNN, \GGF and \VBF. In the resonant search and the anomalous couplings interpretation only the \GGF process is considered, therefore there is only one signal category.
To form the background categories, the DNN classes for minor background processes are grouped together with a major background with a similar topology, in order to simplify the fitting process and reduce the statistical uncertainties.
In the dilepton channel,  multiboson events are categorised together with the DY events to form the ``DY + Multiboson'' category, while single top quark, \ttZ, SM single Higgs boson processes and others are included in the same category as the dominant \ttbar background forming the ``Top + Other'' category.
In the single-lepton channel, single top quark and SM single Higgs boson processes are included in the same event category as the dominant \ttbar background, ``Top + Higgs'' category, while all other processes are included in the same category as the $\PW{+}\text{jets}$ background, ``$\PW{+}\text{jets}$ + Other''.

\begin{table}[htb!]
  \centering
  \topcaption{Summary of the categories of events according to the DNN-based multiclassification and \Hbb topology for the single-lepton channel. The VBF category is considered only in the nonresonant search.}
  \begin{tabular}{cccc}
    \multicolumn{4}{c}{Single-lepton channel}  \\ \hline
    Categories                  & \multicolumn{3}{c}{Subcategories}                         \\ \hline
    \HH(\GGF)                   & Resolved 1b                       & Resolved 2b & Boosted \\
    \HH(\VBF)                   & Resolved 1b                       & Resolved 2b & Boosted \\
    Top + Higgs                 & \multicolumn{2}{c}{Resolved}      & Boosted               \\
    $\PW{+}\text{jets}$ + Other & \multicolumn{3}{c}{Inclusive}                             
  \end{tabular}
  \label{tab:Strategy_scheme_SL}
\end{table}

\begin{table}[htb!]
  \centering
  \topcaption{Summary of the categories of events according to the DNN-based multiclassification and \Hbb topology for the dilepton channel. The VBF category is considered only in the nonresonant search.}
  \begin{tabular}{cccc}
    \multicolumn{4}{c}{Dilepton channel}\\ \hline
    Categories      & \multicolumn{3}{c}{Subcategories}                         \\ \hline
    \HH(\GGF)       & Resolved 1b                       & Resolved 2b & Boosted \\
    \HH(\VBF)       & Resolved 1b                       & Resolved 2b & Boosted \\
    Top + Other     & \multicolumn{2}{c}{Resolved}      & Boosted               \\
    DY + Multiboson & \multicolumn{3}{c}{Inclusive}                             
  \end{tabular}
  \label{tab:Strategy_scheme_DL}
\end{table}

The signal categories are further divided into subcategories according to the \Pbottom jet topology and multiplicity.
Events with one large-radius jet,  as defined in Section~\ref{sec:ak8jets}, are considered in the boosted category if they are \Pbottom-tagged.
If there are large-radius jets that are not \Pbottom-tagged or if there is no large-radius jet at all, the events are divided according to the number of small-radius jets into events with exactly 1 \Pbottom-tagged jet and events with at least 2 \Pbottom-tagged jets.
Remaining events without any \Pbottom-tagged jet will be considered for the data-driven estimation of the DY background, as described in Section~\ref{sec:backgroundEstimation_DY}.
The categories for each decay channel are summarised in Tables~\ref{tab:Strategy_scheme_SL} and~\ref{tab:Strategy_scheme_DL}.
The total number of categories in both dilepton and single-lepton channels is nine for the nonresonant interpretations and six for the resonant, as we only consider \GGF for the latter.

\begin{figure}[htb!]
  \centering
  \includegraphics[width=0.49\textwidth]{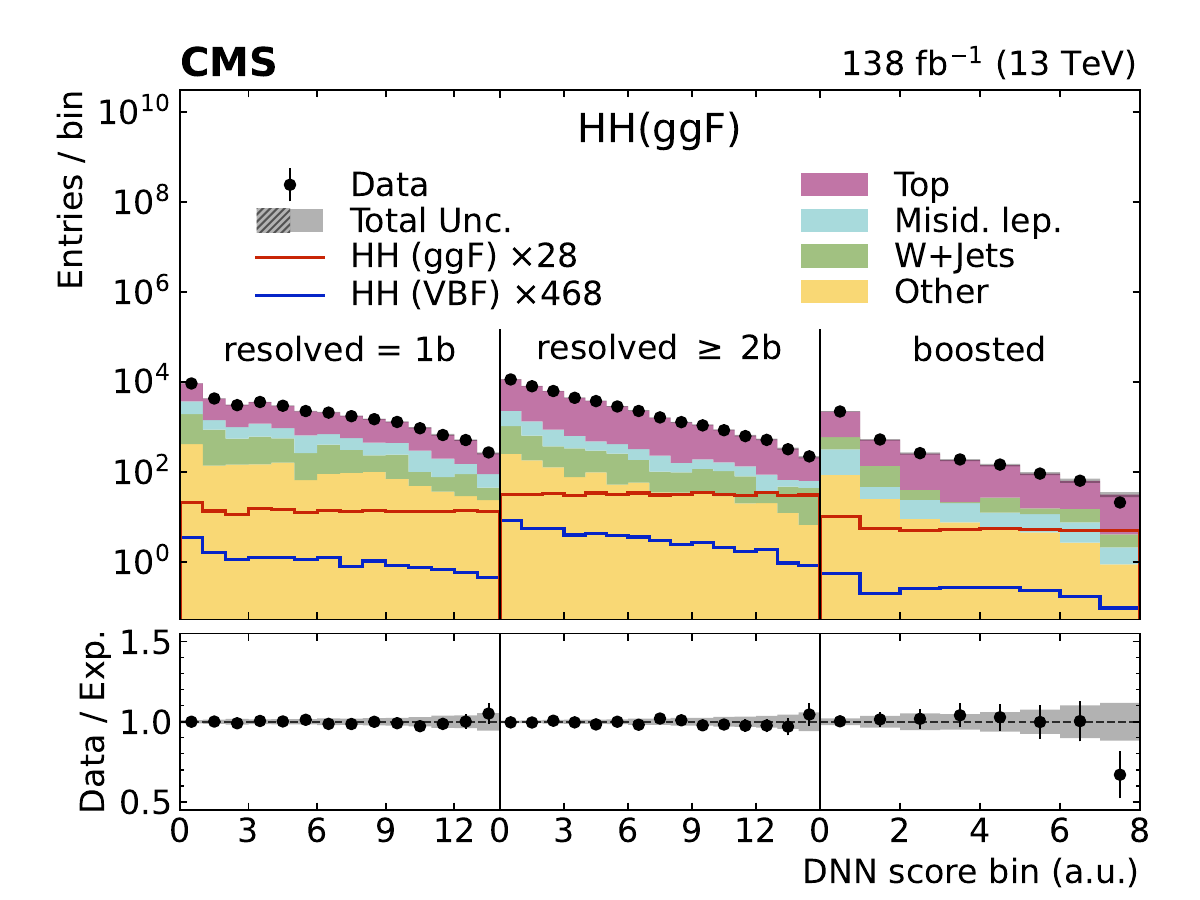}
  \includegraphics[width=0.49\textwidth]{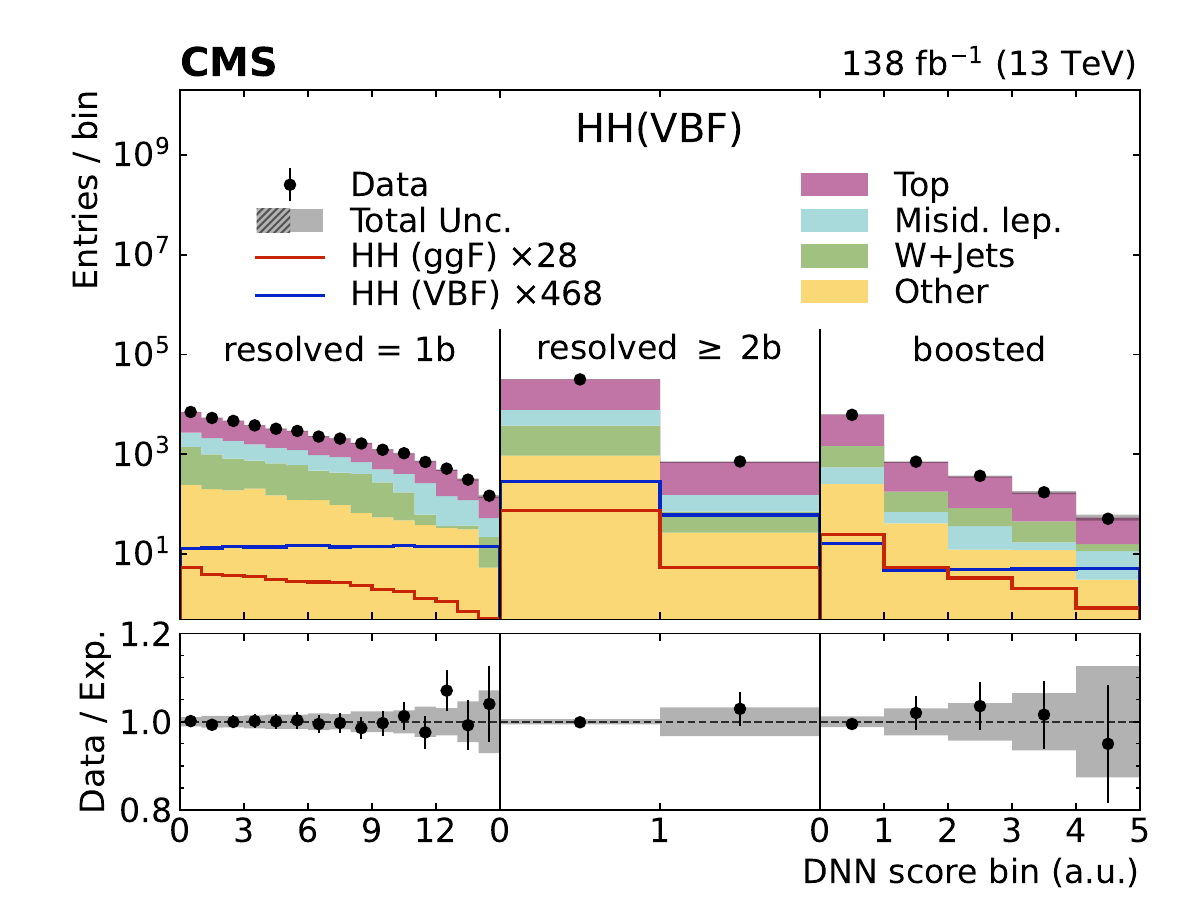}\\
  \includegraphics[width=0.49\textwidth]{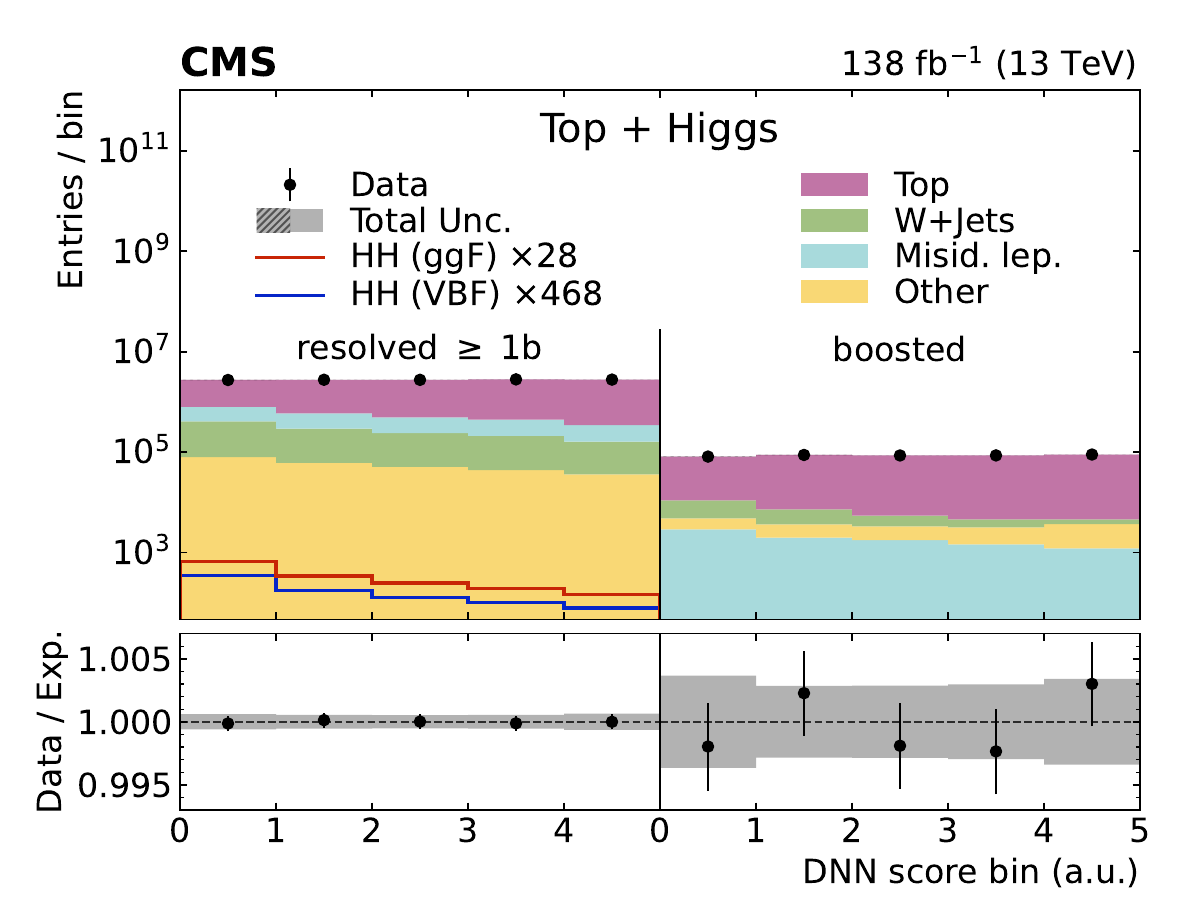}
  \includegraphics[width=0.49\textwidth]{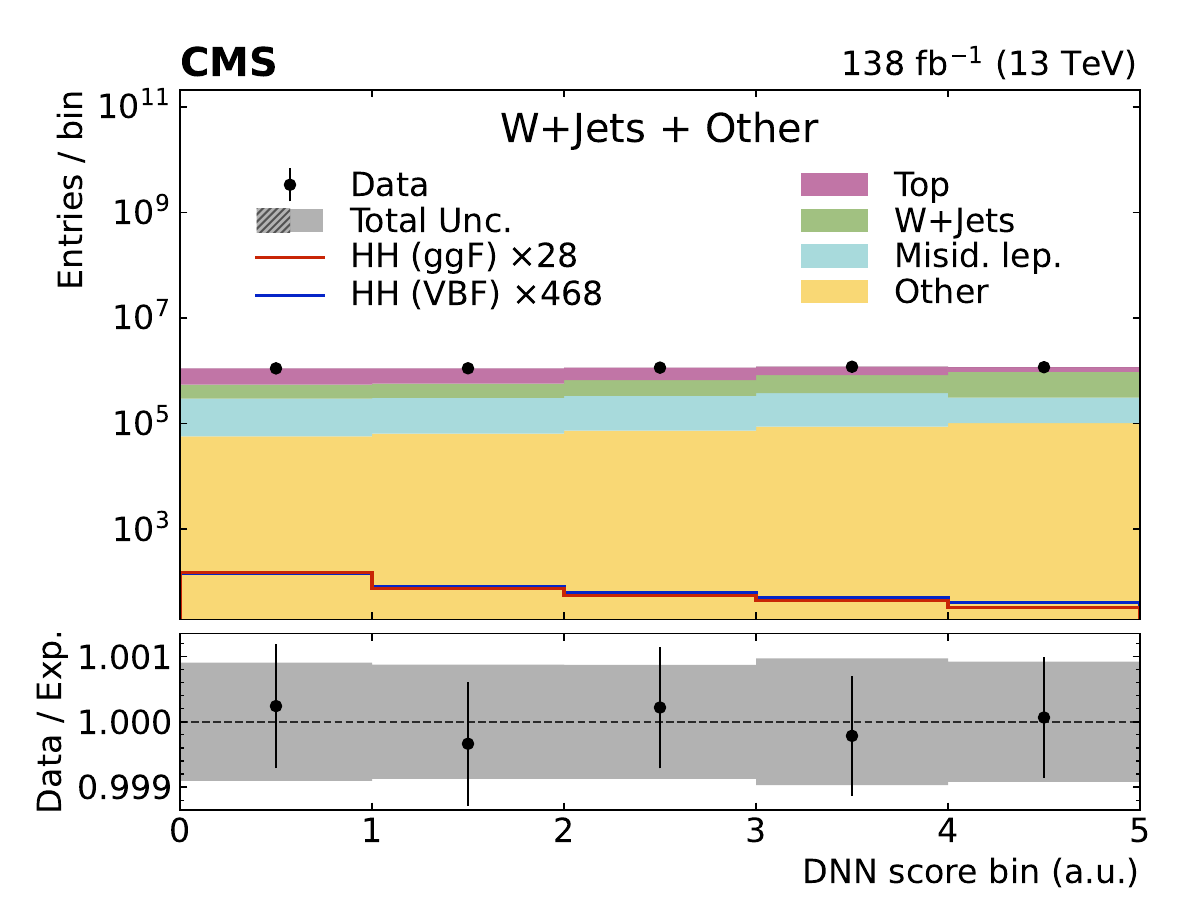}
  \caption{The distributions of the DNN discriminants of the nonresonant search for each event category for the single-lepton channel, after performing a maximum likelihood fit to the same distributions in data. The DNN discriminant for the \HH(\GGF) category is shown on the upper left, \HH(\VBF) on the upper right, Top+Higgs on the lower left and $\PW{+}\text{jets}$ + Other on the lower right. The event categories are summarised in Table~\ref{tab:Strategy_scheme_SL}. The signal shown is scaled to the observed upper limit on its cross section.}
  \label{fig:SL_DNN}
\end{figure}

\begin{figure}[htb!]
  \centering
  \includegraphics[width=0.49\textwidth]{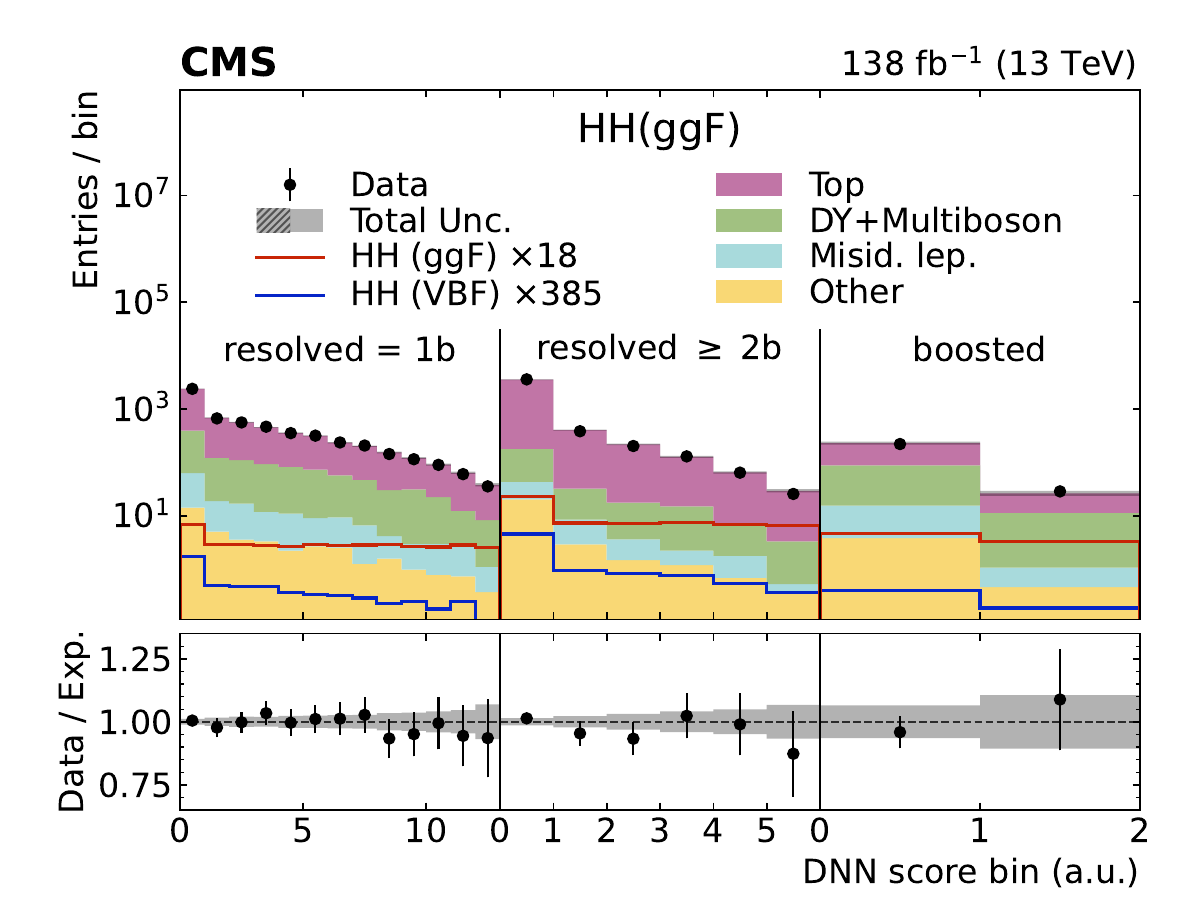}
  \includegraphics[width=0.49\textwidth]{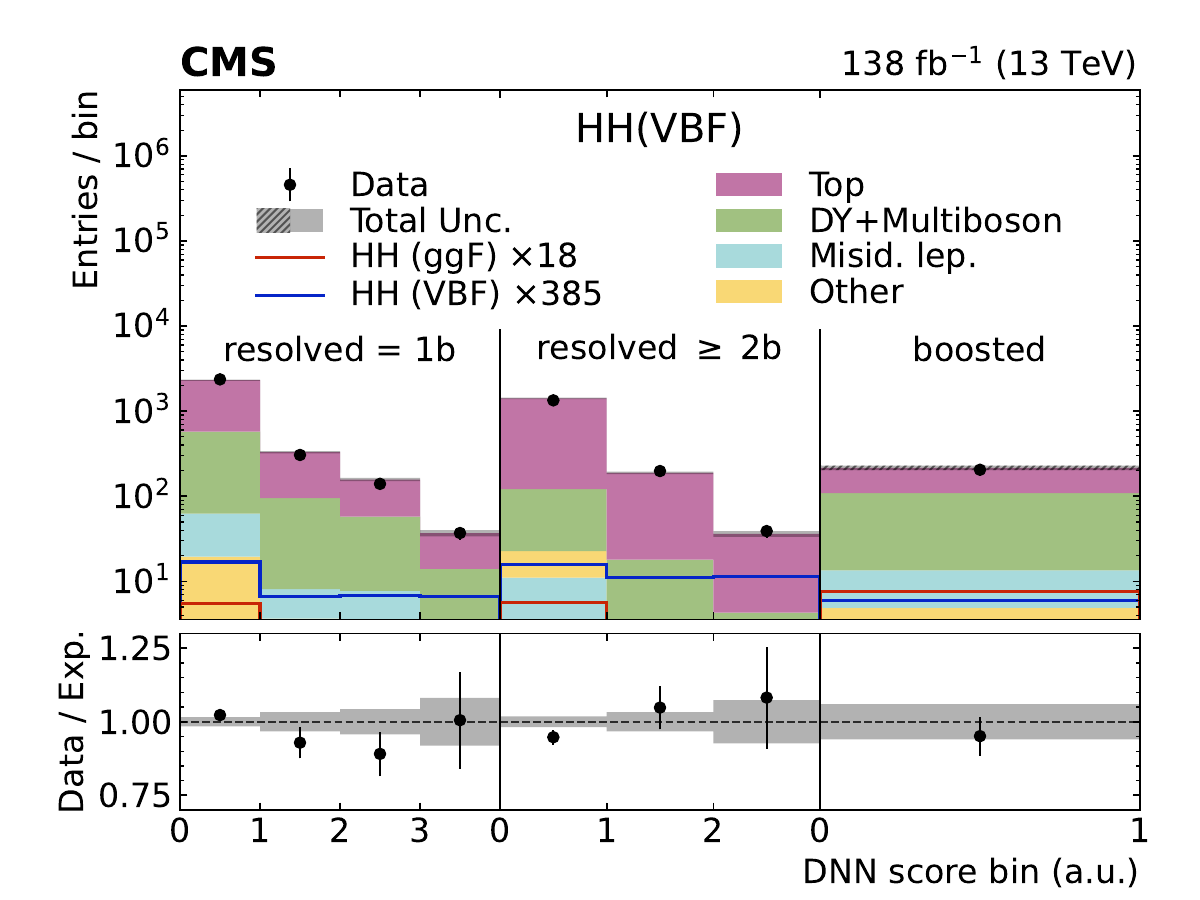}\\
  \includegraphics[width=0.49\textwidth]{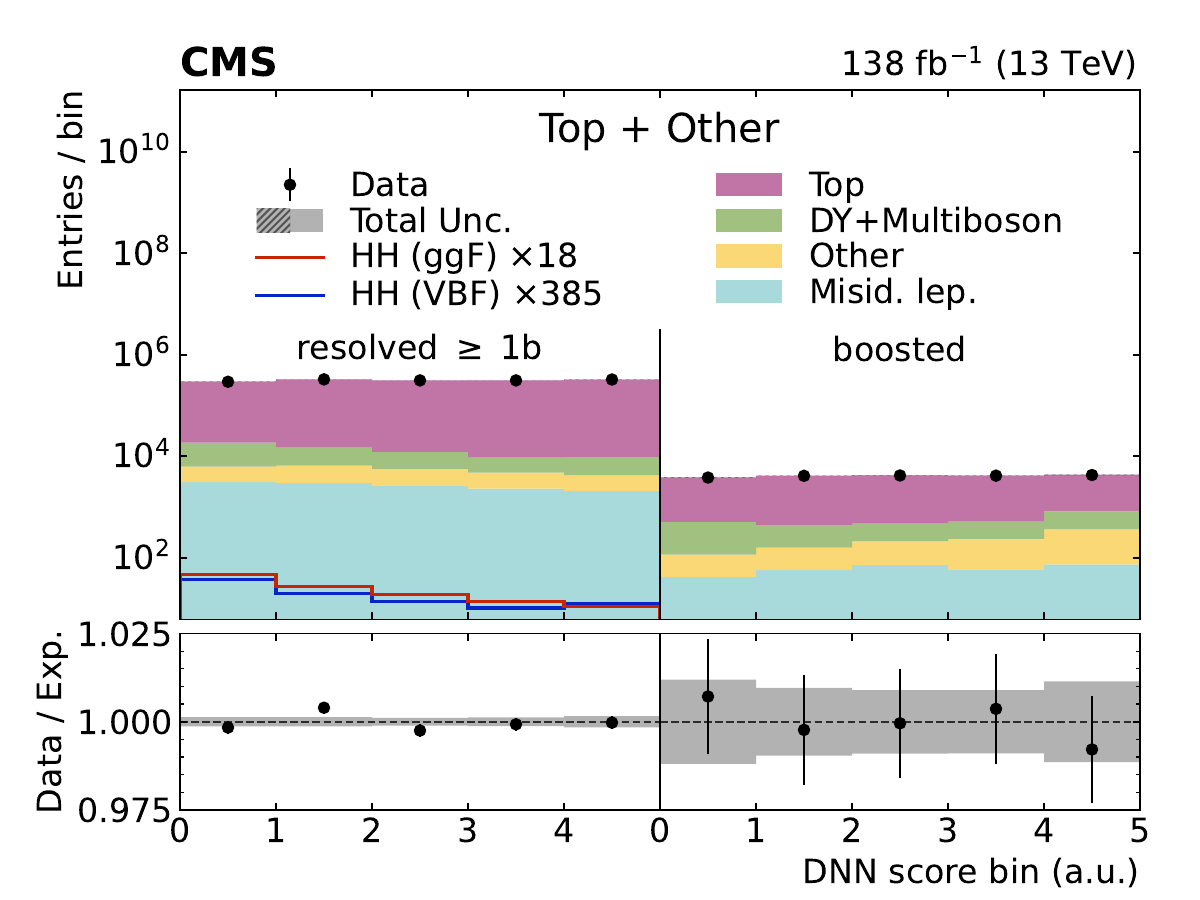}
  \includegraphics[width=0.49\textwidth]{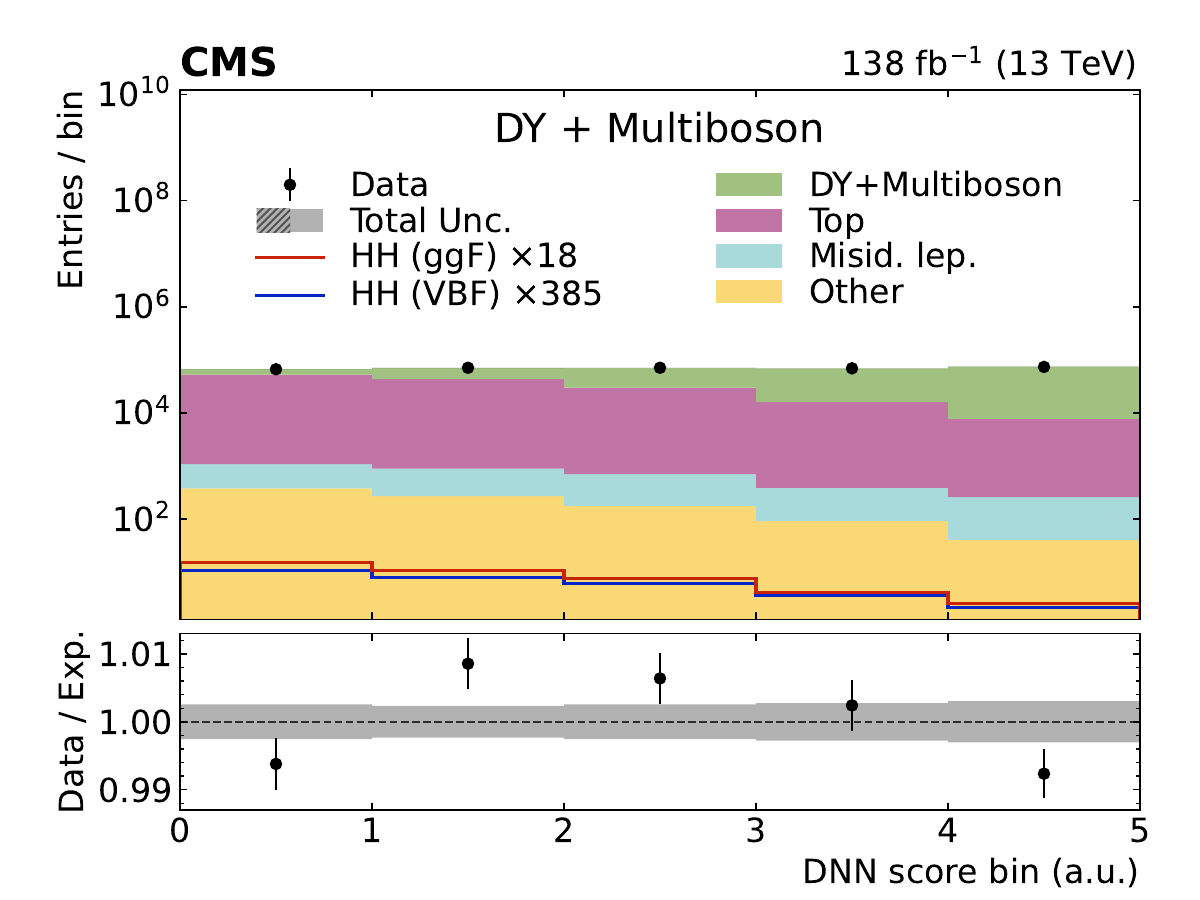}
  \caption{The distributions of the DNN discriminants of the nonresonant search for each event category for the dilepton channel, after performing a maximum likelihood fit to the same distributions in data. The DNN discriminant for the \HH(\GGF) category is shown on the upper left, \HH(\VBF) on the upper right, Top+Other  on the lower left and DY+Multiboson on the lower right. The event categories are summarised in Table~\ref{tab:Strategy_scheme_DL}. The signal shown is scaled to the observed upper limit on its cross section.}
  \label{fig:DL_DNN}
\end{figure}

\begin{figure}[htb!]
  \centering
  \includegraphics[width=0.49\textwidth]{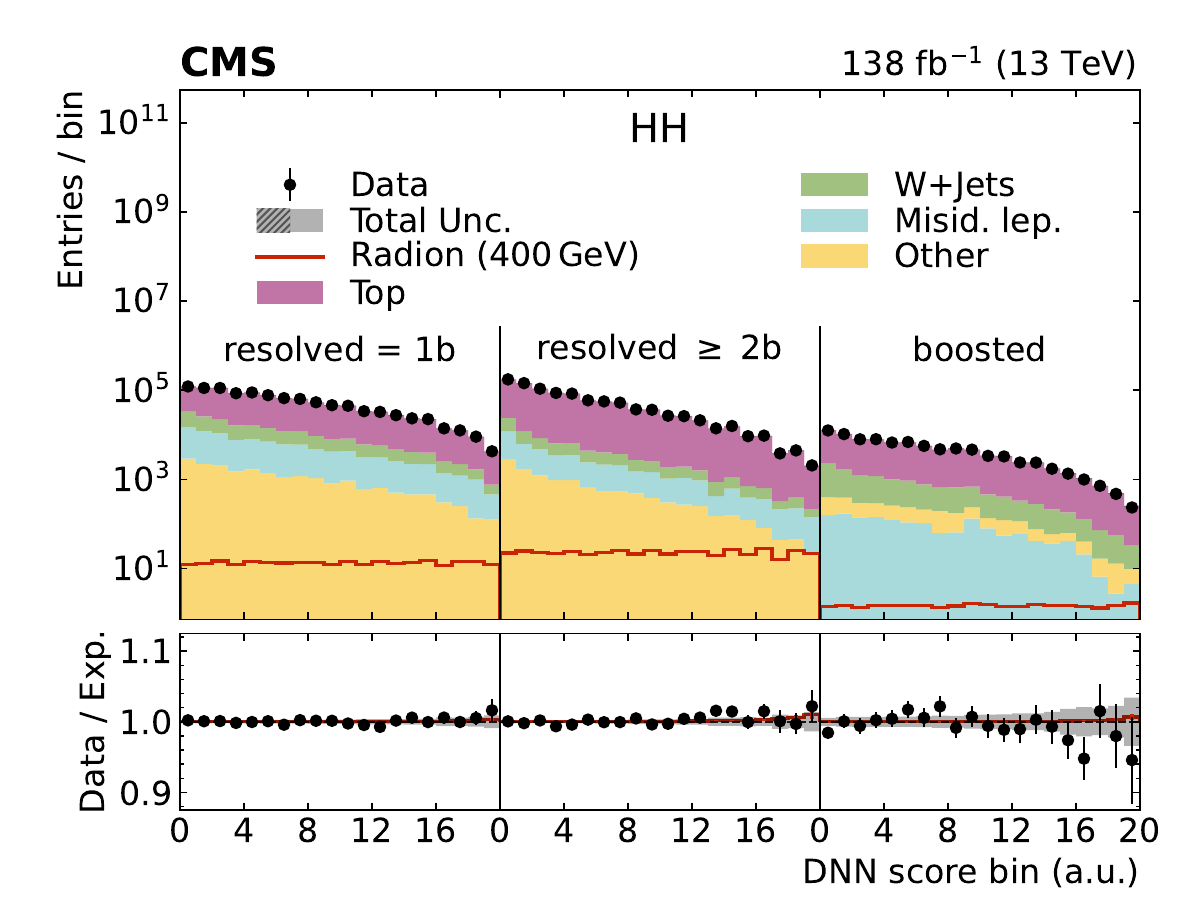}
  \includegraphics[width=0.49\textwidth]{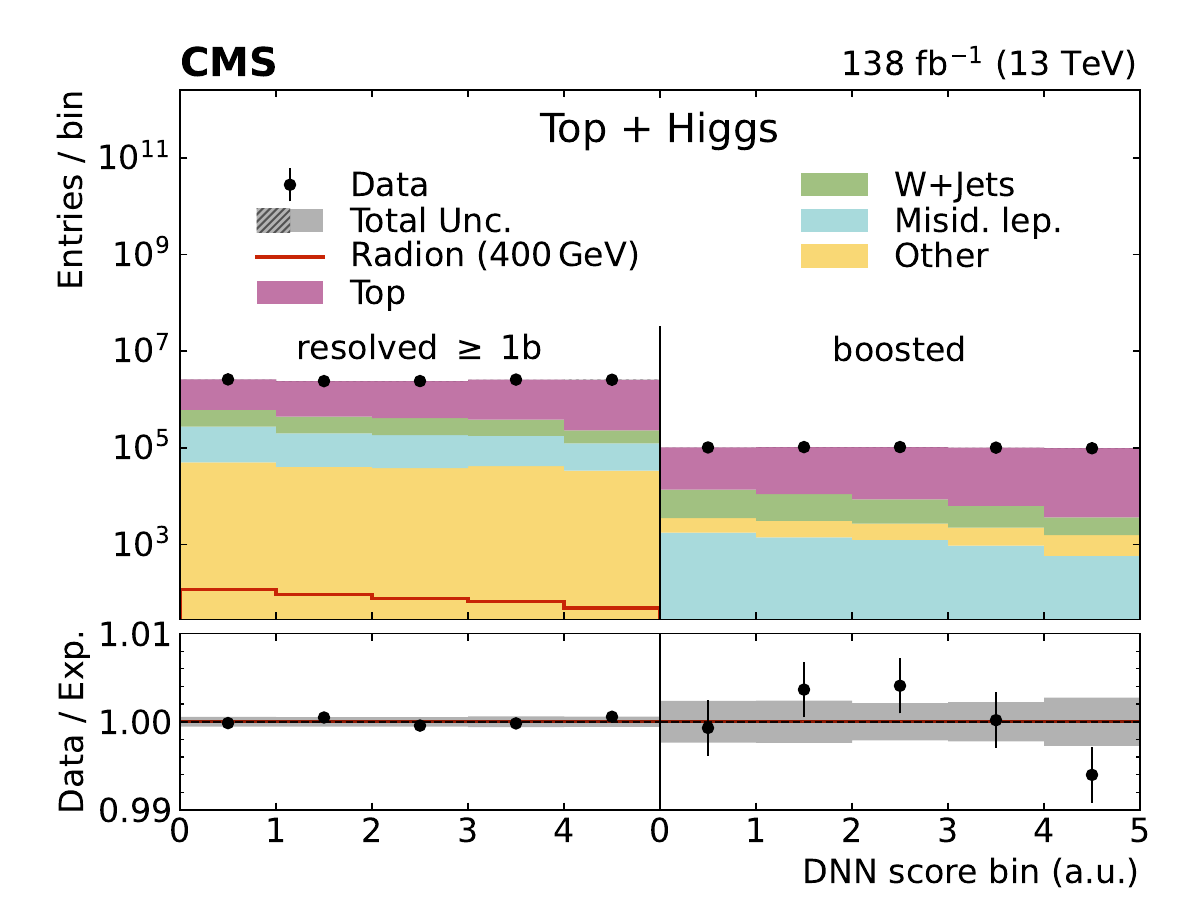}\\
  \includegraphics[width=0.49\textwidth]{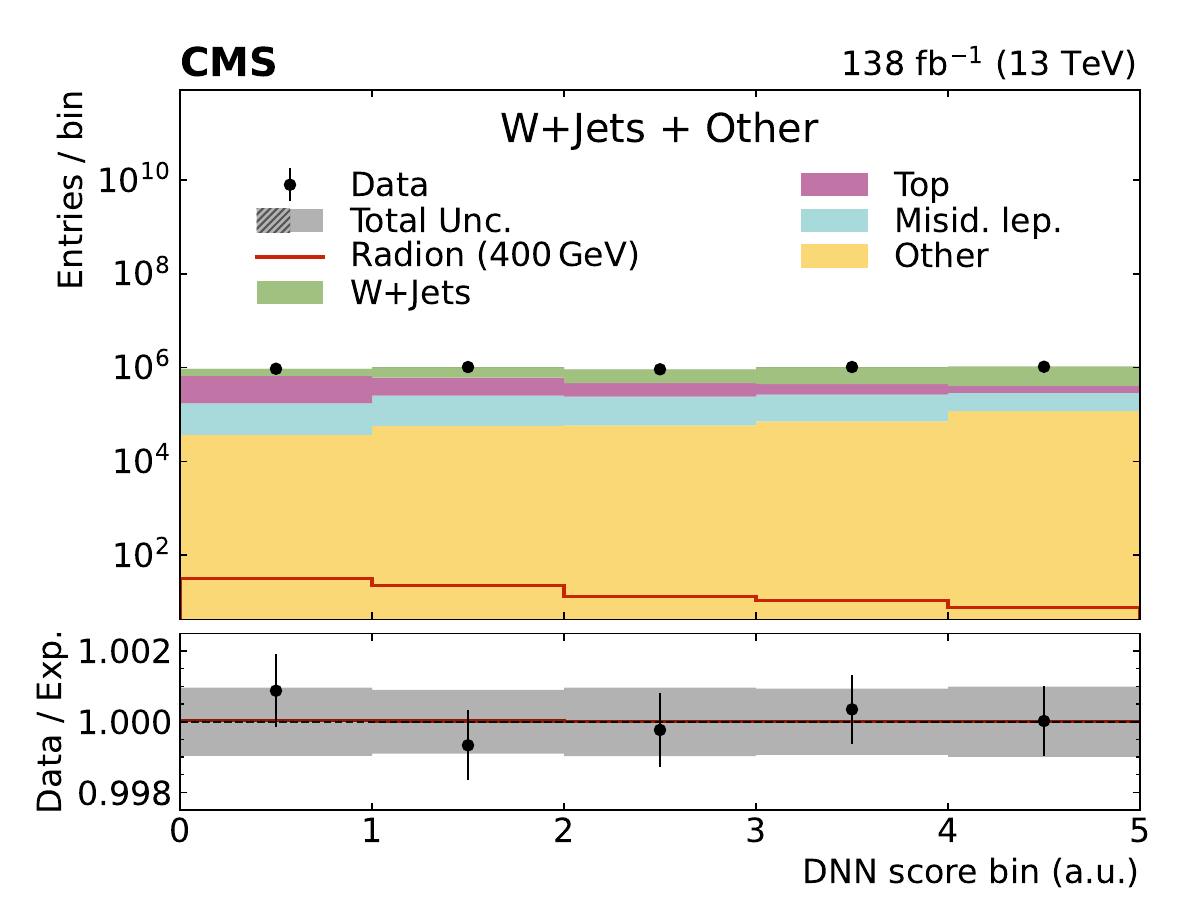}
  \caption{The distributions of the DNN discriminants of the resonant search for each event category for the single-lepton channel, after performing a maximum likelihood fit to the same distributions in data. The DNN shown corresponds to a scalar resonance with mass 400\GeV. The DNN discriminant for the \HH(\GGF) category is shown on the upper left, Top+Higgs on the upper right and $\PW{+}\text{jets}$ + Other on the lower. The event categories are summarised in Table~\ref{tab:Strategy_scheme_SL}. The signal shown is scaled to a cross section of 1\pb.}
  \label{fig:DNNoutput_resonant_SL}
\end{figure}

\begin{figure}[htb!p]
  \centering
  \includegraphics[width=0.49\textwidth]{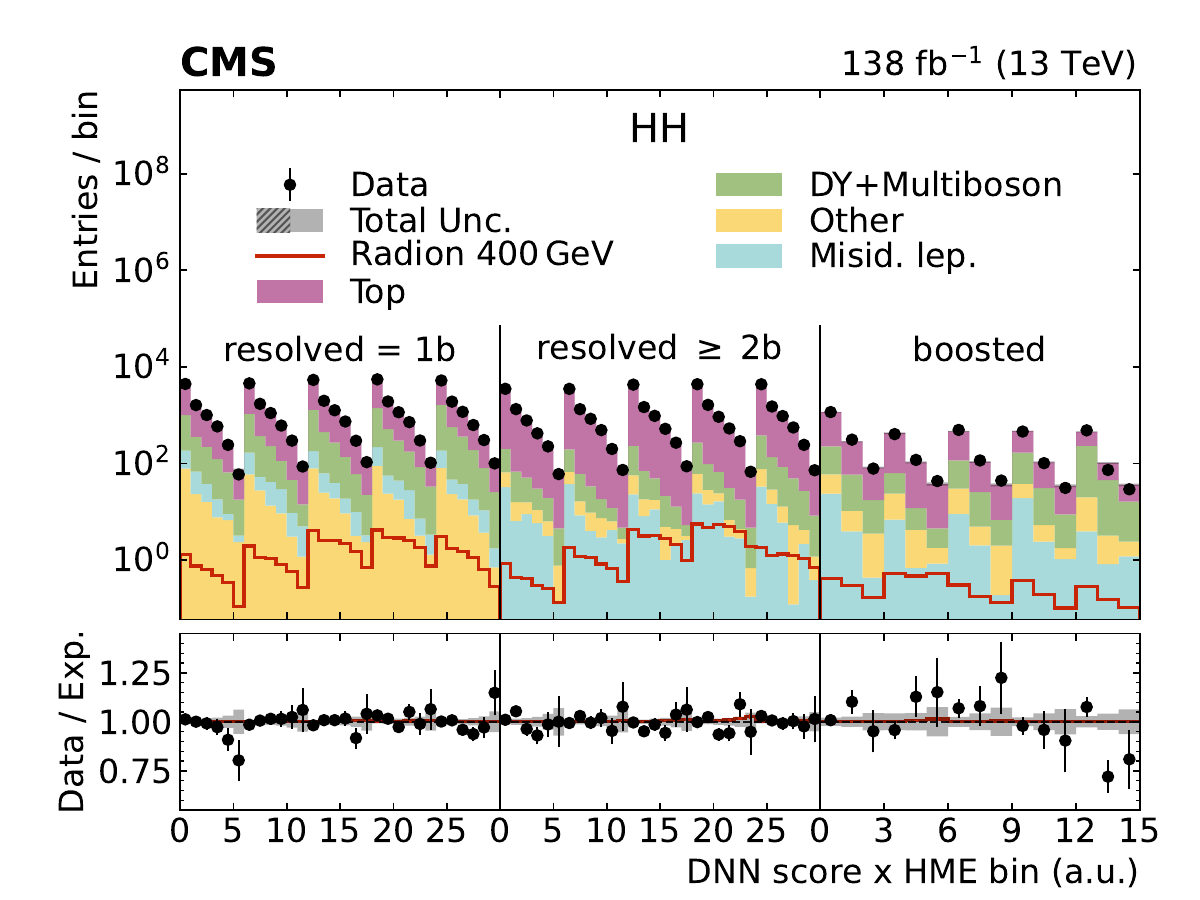}
  \includegraphics[width=0.49\textwidth]{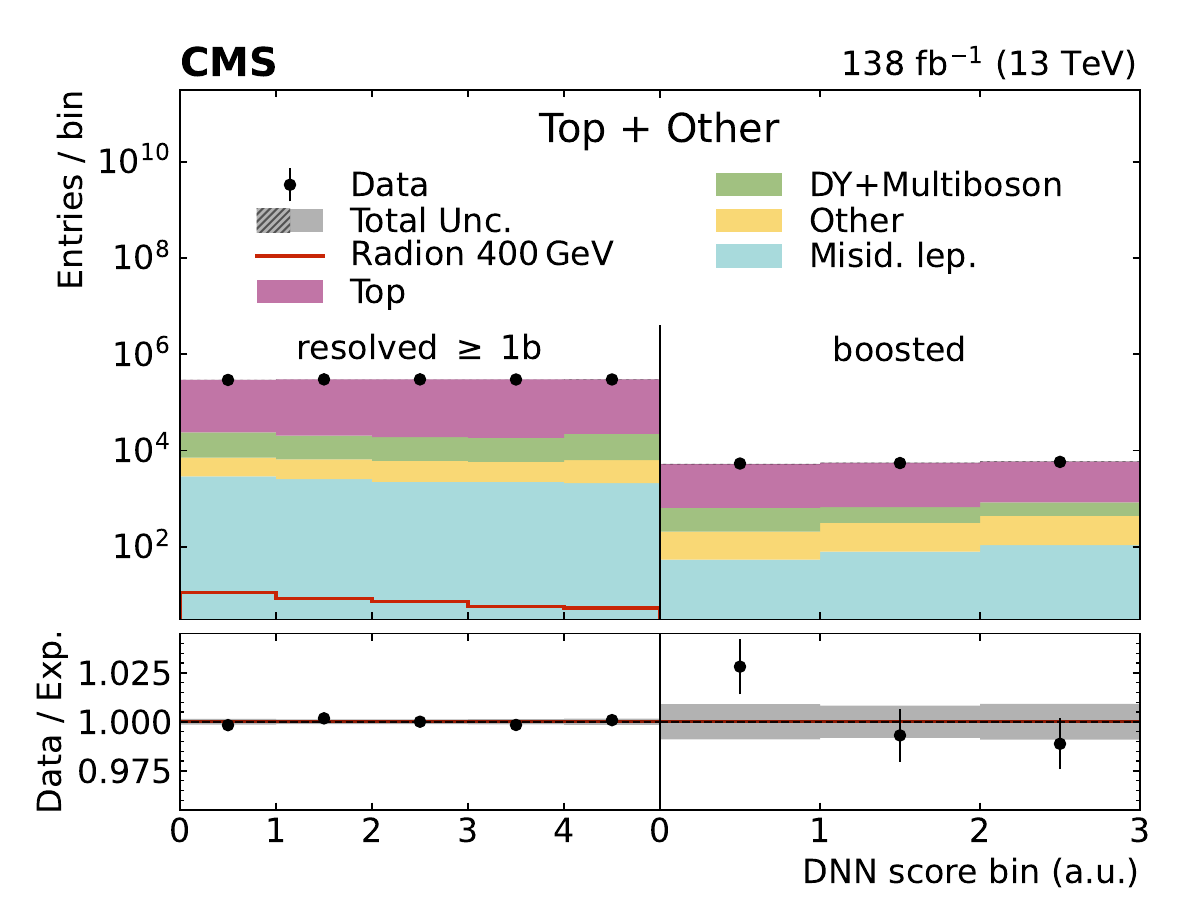}\\
  \includegraphics[width=0.49\textwidth]{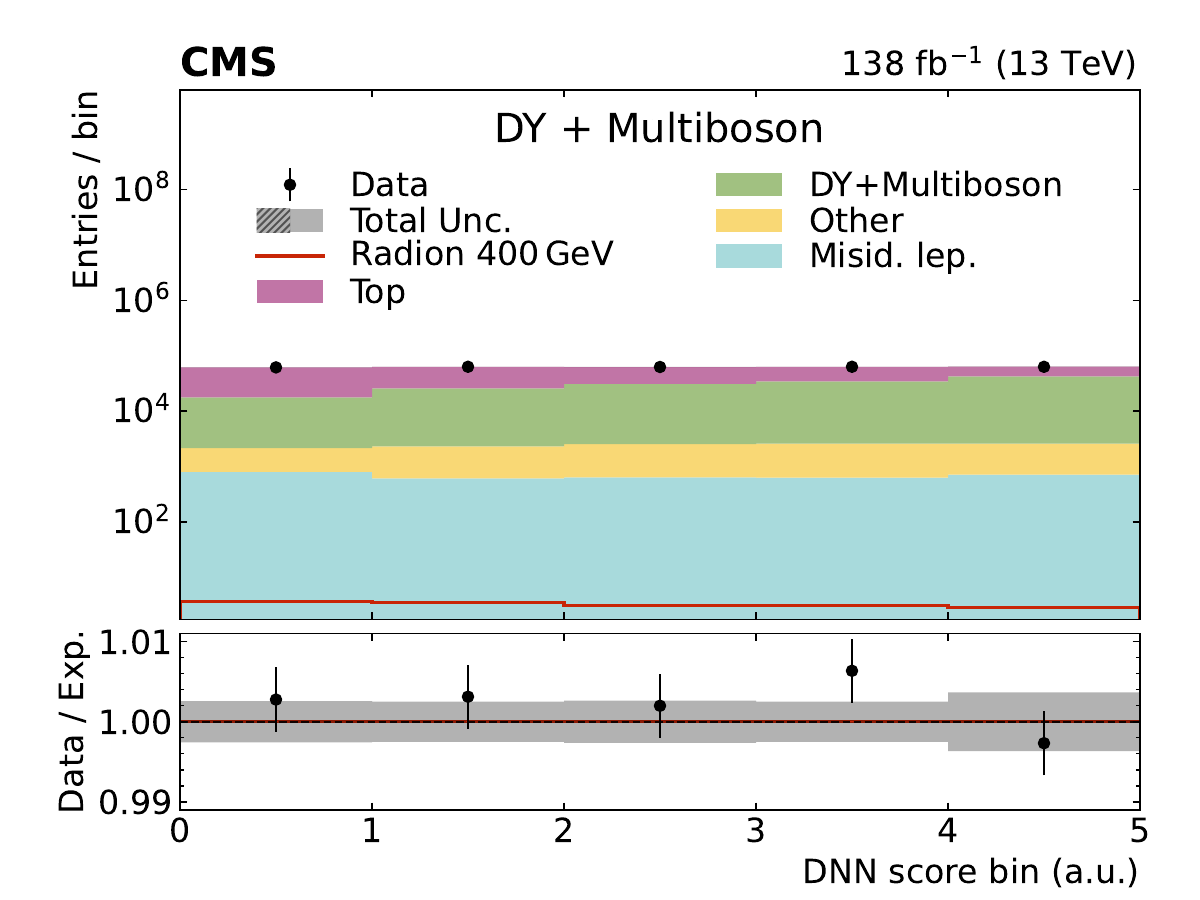}
  \caption{The distributions of the DNN discriminants of the resonant search for each event category for the dilepton channel, after performing a maximum likelihood fit to the same distributions in data. The DNN shown corresponds to a scalar resonance with mass 400\GeV. The DNN discriminant for the \HH(\GGF) category divided in bins of HME is shown on the upper left, Top+Other upper right and DY+Multiboson on the lower. The event categories are summarised in Table~\ref{tab:Strategy_scheme_DL}. The signal shown is scaled to a cross section of 1\pb.}
  \label{fig:DNNoutput_resonant_DL}
\end{figure}

The discriminants used in the maximum likelihood fit for signal extraction are the DNN output scores for each category and channel, combined into a single likelihood function, and are shown in Figs.~\ref{fig:SL_DNN},~\ref{fig:DL_DNN},~\ref{fig:DNNoutput_resonant_SL}, and~\ref{fig:DNNoutput_resonant_DL}. The binning for the signal categories is performed using quantile binning such that the signal response is flat, while the reverse is done for the background categories. The optimisation of the binning was done separately for each data taking year, while the figures contain all the data. Therefore for illustration purposes the binning shown might have slight differences to what was used in the signal extraction and the signal response may not appear flat. One exception is the resonant search in the dilepton channel in which the DNN score of the signal categories is split in bins of the Heavy Mass Estimator (HME)~\cite{PhysRevD.96.035007} as shown in Fig.~\ref{fig:DNNoutput_resonant_DL}.
This variable estimates the most likely invariant mass of the heavy resonance, considering the two neutrinos from the $\PW$ bosons leptonic decays in the final state. In this case, the maximum likelihood fit is performed on the two-dimensional (2D) distribution of the DNN output score and the HME.

\section{Background estimation}\label{Background_estimation}

The shape of the \ttbar contribution for both channels is estimated using simulated events.
The \ttbar and single top quark normalisations are determined by the maximum likelihood fit, and they are constrained by the ``Top+Higgs'' or ``Top+Other" event category, depending on the channel.
The same procedure is used for the $\PW{+}\text{jets}$ background in the single-lepton channel.
To estimate the background contribution of jets misidentified as leptons and, in the dilepton channel, the DY background contribution, control samples in data are used.
The rest of the backgrounds are estimated using simulated events.

\subsection{Estimation of the misidentified-lepton background}
\label{sec:backgroundEstimation_fakes}

The background arising from events with misidentified leptons is estimated using the ``fake-factor'' method in Refs.~\cite{ttHmultilepton} and~\cite{HIG17018}.
A sample of events is selected by requiring that all lepton criteria of Section \ref{Event_selection} must be satisfied, with the exception that at least one electron or muon passes the medium and fails the tight selection (with \pt requirements replaced by the same ones on $\pt^{\text{cone}}$), preventing overlap with the signal region.

An estimate of the misidentified-lepton background in the signal region is obtained by applying suitably chosen weights to these events.
The weights, denoted by the symbol $w$, are given by the expression:
  \begin{equation}
    w = (-1)^{n+1} \, \prod_{i=1}^{n} \, \frac{f_{i}}{1 - f_{i}},
    \label{eq:FF_weights}
  \end{equation}
where $f_{i}$ denotes the probability for an electron or a muon that passes the medium selection to also satisfy the tight selection. The probabilities $f_{i}$ are measured separately for electrons and muons, as described in Ref.~\cite{ttHmultilepton} in a control region dominated by the multijet background.

The product extends over the number of electrons and muons that pass the medium and fail the tight selection ($n$). In case an event of the single-lepton (dilepton) channel contains more than one (two) medium leptons, only the leading (and subleading) lepton in $\pt^{\text{cone}}$ is considered when computing the weights according to Eq.~(\ref{eq:FF_weights}), and thus $n = 1$ ($n = 1$ or 2) in the single-lepton (dilepton) channel. For $n = 2$ the weights become negative to avoid double counting. The contributions of other backgrounds are subtracted based on the expectation from simulation.

\subsection{Estimation of the Drell--Yan background}
\label{sec:backgroundEstimation_DY}
The DY background in the dilepton channel is estimated using data events that pass the nominal event selection but have no reconstructed \Pbottom-tagged jets.
This ``0 \Pbottom tag region'' is found to be dominated by DY events.
We create additional DY enriched regions by inverting the selection requirement on the dilepton mass targeting DY events.
This inverse selection is applied for events with 0, 1 and 2 \Pbottom-tagged jets.
Using events with dilepton mass within $10$\GeV from the Z boson mass, we calculate transfer weights from the 0 \Pbottom tag region to the 1 and 2 \Pbottom tag categories.
In the signal region, these weights are applied to the 0 \Pbottom tag region using two orthogonal sets of events to estimate the DY in the 1 and 2 \Pbottom tag regions, in order to avoid bias from using the same events twice.
The same process is applied to the boosted categories.
The weights are found to be independent of the lepton flavour, and the proportion of DY events in the $\Pe\PGm$ channel from tau lepton decays to \Pe or \PGm is insignificant.
Therefore we calculate the weights in $\Pe\Pe$ and $\PGm\PGm$ events simultaneously and apply them to the $\Pe\Pe$, $\PGm\PGm$ and $\Pe\PGm$ events.
When extracting the shape of the DY distribution in the 0 \Pbottom tag region from the data, the contribution from other background sources is estimated using simulated events.
\section{Systematic uncertainties}
\label{systematics}
Systematic uncertainties are introduced as nuisance parameters in the maximum likelihood fit used to extract the signal.

A number of systematic uncertainties are considered that affect the yield and the shapes of the \HH signal and the background processes.
Theoretical uncertainties in the strong interaction coupling $\alpha_s$, and parton distribution function shapes that affect the cross section of all the simulated processes are included.

Theoretical uncertainties in the nonresonant \HH cross section via \GGF are applied as a function of \klambda
and include renormalisation and factorisation scale uncertainties, including the mass scale of the top quark~\cite{Baglio:2020wgt}. In the SM this uncertainty amounts to $+$6\%/$-$26\%. An additional factor of $\pm$3.0\% is applied to account for PDF$+\alpha_s$ uncertainties. 
The uncertainties on the \VBF production cross section include a $+$0.03/$-$0.04\% (scale) and a $\pm$2.1\% (PDF$+\alpha_s$) ~\cite{Ling:2014sne, Dreyer:2018rfu} uncertainties. All uncertainties on the signal cross section are only applied when quoting limits relative to the theoretically predicted cross section.
An additional 10\% normalisation uncertainty is applied to the \VBF, related to the colour correlated recoil scheme~\cite{dipoleShoweStudies} used in \PYTHIA. It is estimated by comparing the nominal simulated samples produced with the default global recoil scheme to samples simulated with the dipole-recoil scheme.

The uncertainty in the top quark mass value assumed in the simulations of the \ttbar background is derived by varying the mass value by $-$2.7/$+$2.8\% ~\cite{deFlorian:2017qfk}. Theoretical uncertainties on the branching fractions~\cite{deFlorian:2016spz} of $\PHiggs$ decays are applied to \HH signal and single Higgs boson background ($\PHiggs\to\bb$, $\PHiggs\to\WW$ and $\PHiggs\to\tau\tau$ branching fractions, $+$1.24/$-$1.26\%, $+$1.53/$-$1.52\%, and $+$1.65/$-$1.63\% respectively).

Other theory uncertainties include electroweak corrections for the $\ttbar\PZ$ ($+$0.0/$-$0.2\%) and $\ttbar\PW$ ($+$0.0/$-$3.2\%) processes as well as PDF weights ($\pm$4.2\% for \ttbar, $\pm$1.2\% for single top quark, $\pm$4.6\% for $\PV\PV$, $\pm$2.8\% for $\ttbar\PZ$, and $\pm$2\% for $\ttbar\PW$). Parton shower acceptance uncertainties are applied as shape variations for all background processes.

A shape uncertainty that corresponds to the NNLO correction of the top quark \pt is applied to the \ttbar simulated samples. The relative uncertainty is 100\%, \ie an uncertainty as large as the correction is applied to the simulated \ttbar events.

The integrated luminosities for the 2016--2018 data taking years separately have 1.2--2.5\% individual uncertainties~\cite{LUM-17-003,LUM-17-004,LUM-18-002}, while the overall uncertainty for the 2016--2018 period is 1.6\%.

During the 2016--2017 data taking, a gradual shift in the timing of the inputs of the ECAL Level-1 trigger in the region at $\abs{\eta} > 2.0$ caused a specific trigger inefficiency. For events containing an electron or a jet with $\pt> 50\GeV$ or $100\GeV$ respectively, in the region $2.5 < \abs{\eta} < 3.0$ the efficiency loss is 10--20\%, depending on \pt, $\eta$, and data taking period. Correction factors are computed from data and applied to the acceptance evaluated from simulation. In addition, a normalisation uncertainty is included in the statistical fit.

A shape uncertainty related to the pileup in simulation and its 5\% \cite{CMS:2018mlc} inelastic pp cross section uncertainty is applied to all simulated samples.

The trigger selection efficiency is corrected to account for differences between data and simulation.
The uncertainties in these corrections applied to simulated samples constitute a shape uncertainty. Similarly, the muon and electron identification efficiencies are also corrected and corresponding shape uncertainties are introduced.

An uncertainty in the efficiency of the selection rejecting jets originating from pileup, as well as uncertainties on jet energy scale and resolution, are used as shape uncertainties for all simulated samples.

Corrections based on the area and energy density of the jet are applied in order to compensate for pileup effects.
Nuisance parameters that affect the shape related to the identification of the jet flavour are also included. Different types of jet flavour contamination are treated with separate parameters.

A set of systematic uncertainties is included in the fit regarding the estimation of the misidentified-lepton background from control samples in data, as well as the DY background.
The uncertainties regarding the backgrounds derived from data are assigned and validated using simulated events through closure tests on the DNN distributions entering the fit to ensure that they cover the observed amount of nonclosure.
For the misidentified-lepton background estimation, they are shape uncertainties in the single-lepton channel and normalisation uncertainties in the dilepton channel.
The DY estimation closure uncertainty, applicable only to the dilepton channel, is a shape uncertainty for all background categories and a normalisation uncertainty for the \HH signal categories.

\section{Results}\label{final_results}
For each signal model considered, a profile binned likelihood fit~\cite{CLSA, CLS1, CLS2} is performed to the distributions of the DNN discriminants for each event category (Figs.~\ref{fig:SL_DNN} to \ref{fig:DNNoutput_resonant_DL}) simultaneously.
For the SM-like signal search the fit includes 18 categories while 12 categories are used for the resonant search and the anomalous couplings interpretation. The categories dominated by background events allow for in-situ constraints on the main background processes.
There are three background-dominated categories for the single-lepton channel and three for the dilepton channel.
No significant excess over the background-only hypothesis is observed. Upper limits are set on nonresonant and resonant Higgs boson pair production at 95\% \CL using the modified frequentist \CLs method in the asymptotic approximation.

\begin{figure}[htb!]
  \centering
  \includegraphics[width=0.75\textwidth]{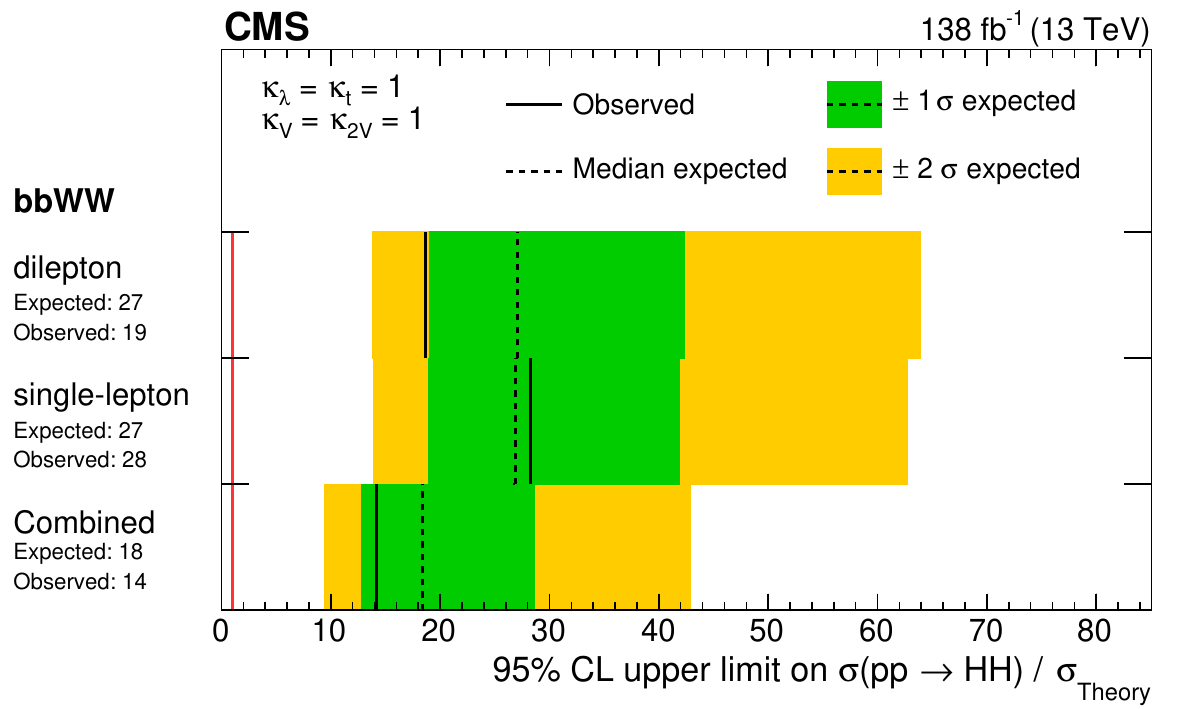}
  \caption{Observed and expected 95\% \CL upper limits on the inclusive nonresonant \HH production cross section divided by the SM prediction obtained for both single-lepton and dilepton channels, and from their combination. The green and yellow bands show the 1 and $2\sigma$ confidence intervals, corresponding to 68 and 95\% \CL , while the red line shows the SM prediction.}
  \label{fig:Limit_SM}
\end{figure}

\begin{figure}[htb!]
  \centering
  \includegraphics[width=0.75\textwidth]{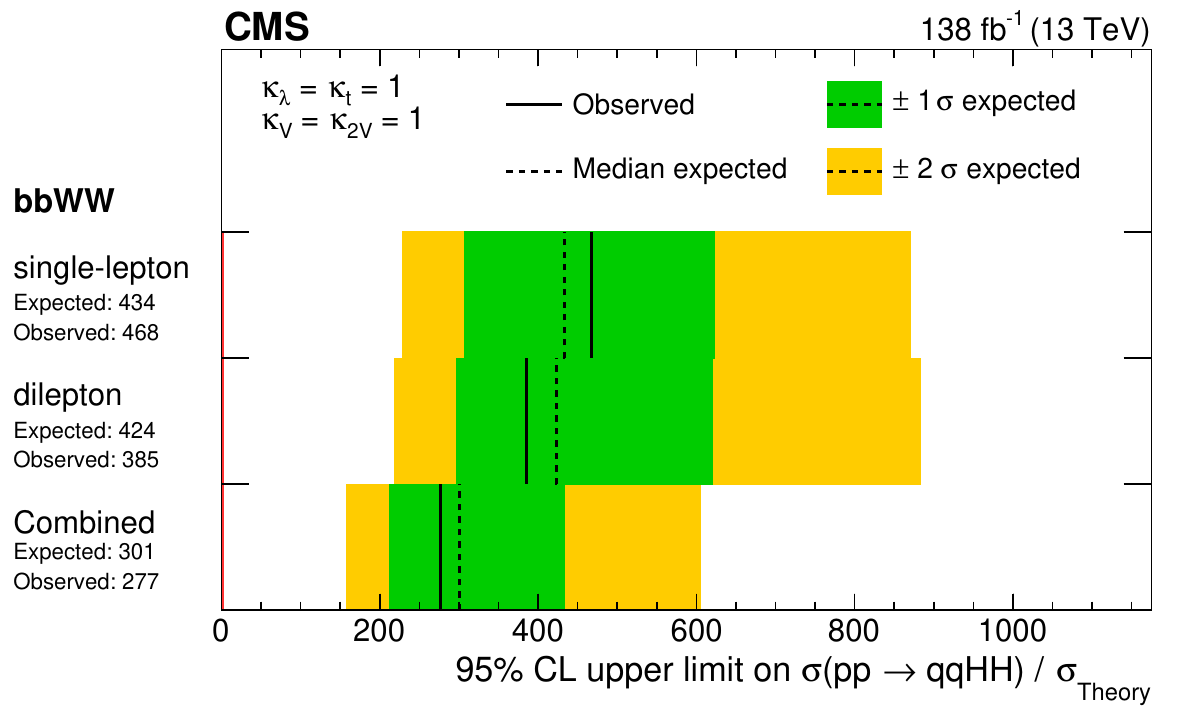}
  \caption{Observed and expected 95\% \CL upper limits on the nonresonant \HH production via vector boson fusion cross section divided by the SM prediction obtained for both single-lepton and dilepton channels, and from their combination. The green and yellow bands show the 1 and $2\sigma$ confidence intervals, corresponding to 68 and 95\% \CL, while the red line shows the SM prediction.}
  \label{fig:Limit_VBF}
\end{figure}

The observed (expected) upper limit on the inclusive $\Pp\Pp\to\HH$ cross section is 14 (18) times the value expected by the SM (Fig.~\ref{fig:Limit_SM}). The observed (expected) limit on the \HH production via \VBF is 277 (301) times the SM value and is shown in Fig.~\ref{fig:Limit_VBF}. In this case the SM value is assumed for the \GGF \HH process. Figs.~\ref{fig:Limit_SM} and ~\ref{fig:Limit_VBF} show the contributions from individual channels as well.

\begin{figure}[htb!]
  \centering
  \includegraphics[width=0.75\textwidth]{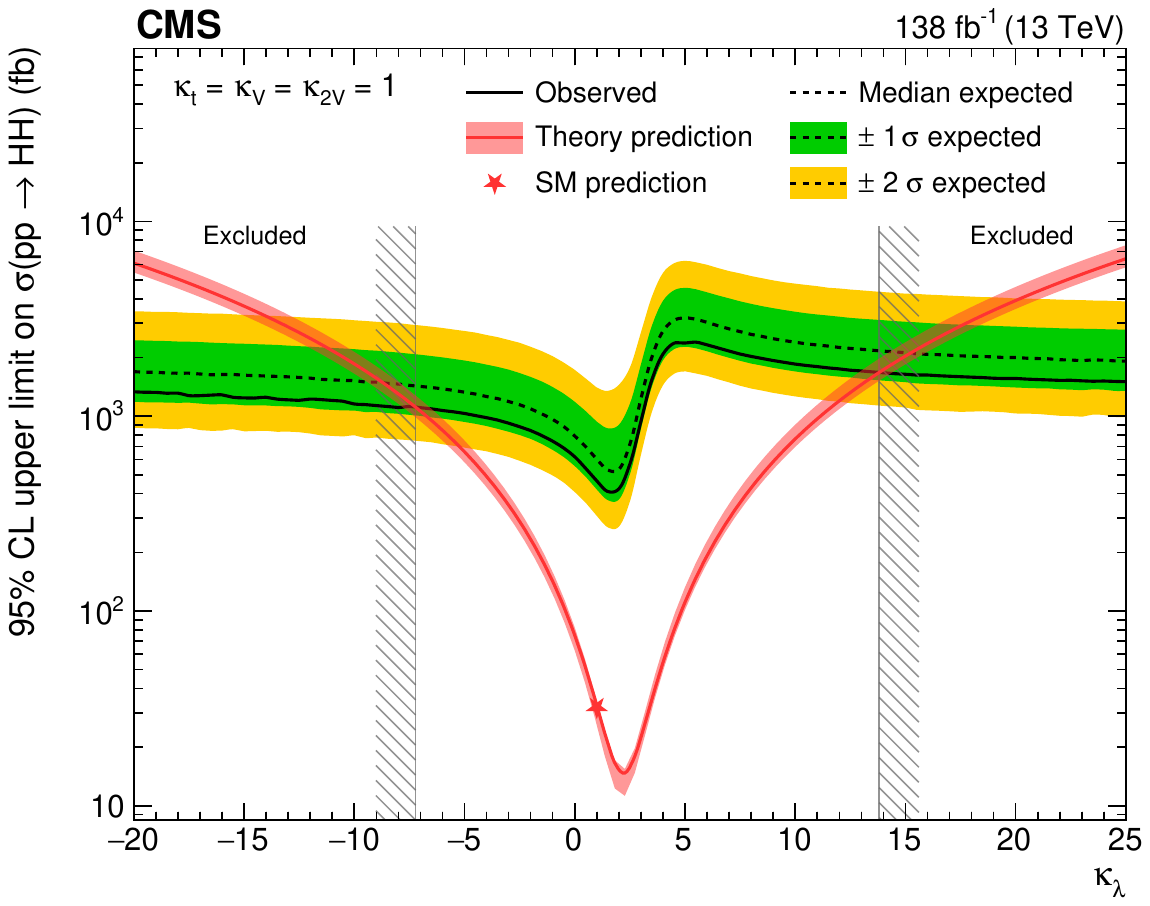}
  \caption{Observed and expected 95\% \CL upper limits on the nonresonant \HH production cross section as a function of the Higgs boson self-coupling strength modifier \klambda. The green and yellow bands show the 1 and $2\sigma$ confidence intervals, corresponding to 68 and 95\% \CL. All Higgs boson couplings other than $\lambda$ are set to the values predicted by the SM. Overlaid in red is the curve representing the predicted \HH production cross section.}
  \label{fig:limit_kl_scan}
\end{figure}

\begin{figure}[htb!]
  \centering
  \includegraphics[width=0.75\textwidth]{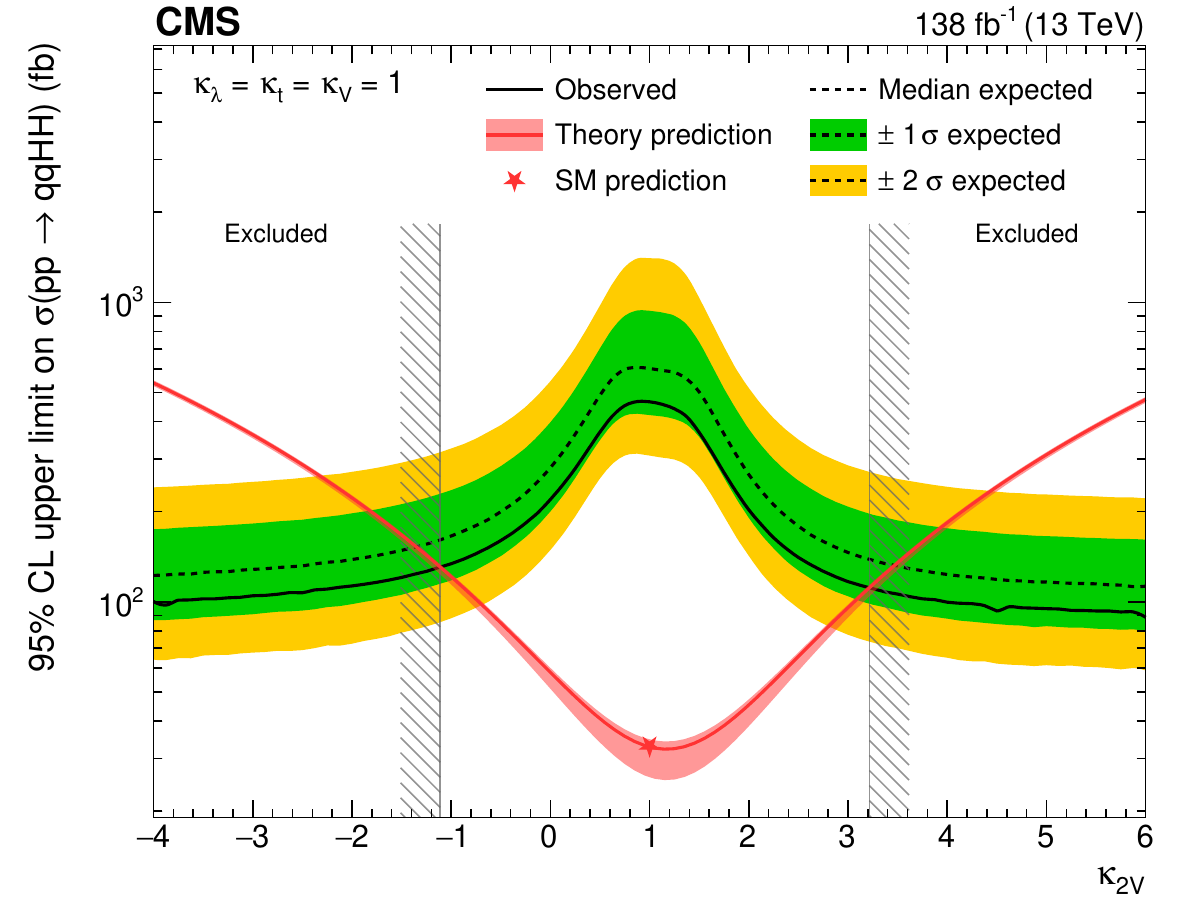}
  \caption{Observed and expected 95\% \CL upper limits on the nonresonant \HH production via \VBF cross section as a function of the effective coupling \CVV. The green and yellow bands show the 1 and  $2\sigma$ confidence intervals, corresponding to 68 and 95\% \CL. The \GGF contribution in this case is set to the SM expectation. All other Higgs boson couplings are set to the values predicted by the SM. Overlaid in red is the curve representing the predicted \HH production cross section.}
  \label{fig:limit_kvv_scan}
\end{figure}

Figures~\ref{fig:limit_kl_scan}~and~\ref{fig:limit_kvv_scan} show the limits on the inclusive and the \VBF production cross section as a function of \klambda and \CVV modifiers, respectively, assuming standard model values for all other couplings.
The \klambda modifier is constrained between $[-7.2,13.8]$ (expected $[-8.7,15.2]$).
The \CVV modifier is constrained between $[-1.1,3.2]$ (expected $[-1.4,3.5]$). The latter result has similar sensitivity to the \bbgg channel which constrained the \CVV modifier in the range $[-1.3,3.5]$ (expected $[-0.9,3.1]$)~\cite{CMS_HHbbgg_R2Legacy}.

\begin{figure}[htb!]
  \centering
  \includegraphics[width=0.75\textwidth]{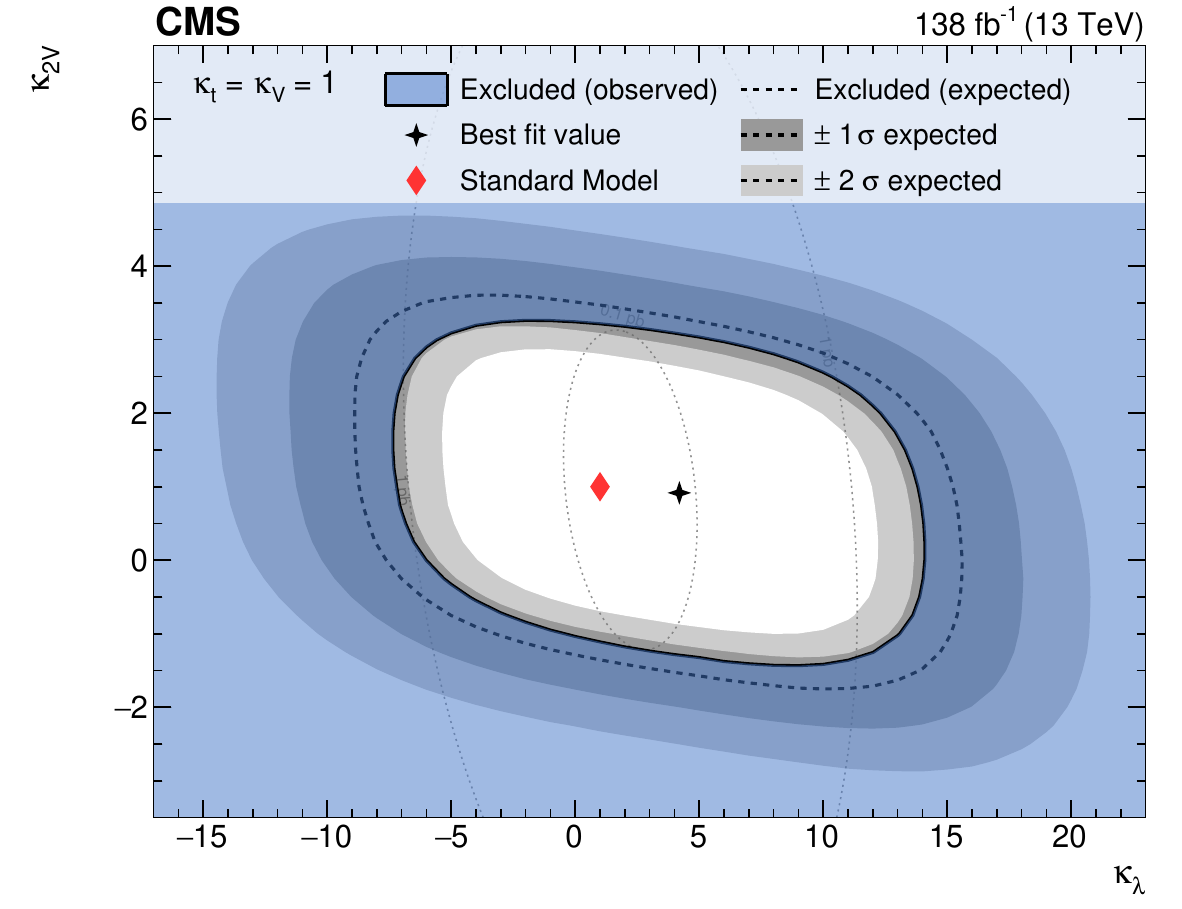}
  \caption{Observed and expected 95\% \CL exclusion limits on the nonresonant \HH production cross section as a function of the effective couplings \klambda and \CVV. The blue area is excluded by the observation. The 1 and $2\sigma$ confidence intervals around the expected median exclusion contour are shown as dark and light-grey areas corresponding to 68 and 95\% respectively. The red diamond shows the SM expectation while the fine dashed lines show the theoretical cross section contours. All other Higgs boson couplings are set to the values predicted by the SM.
  }
  \label{fig:2D_exclusion_kl_kvv}
\end{figure}

\begin{figure}[htb!]
  \centering
  \includegraphics[width=0.75\textwidth]{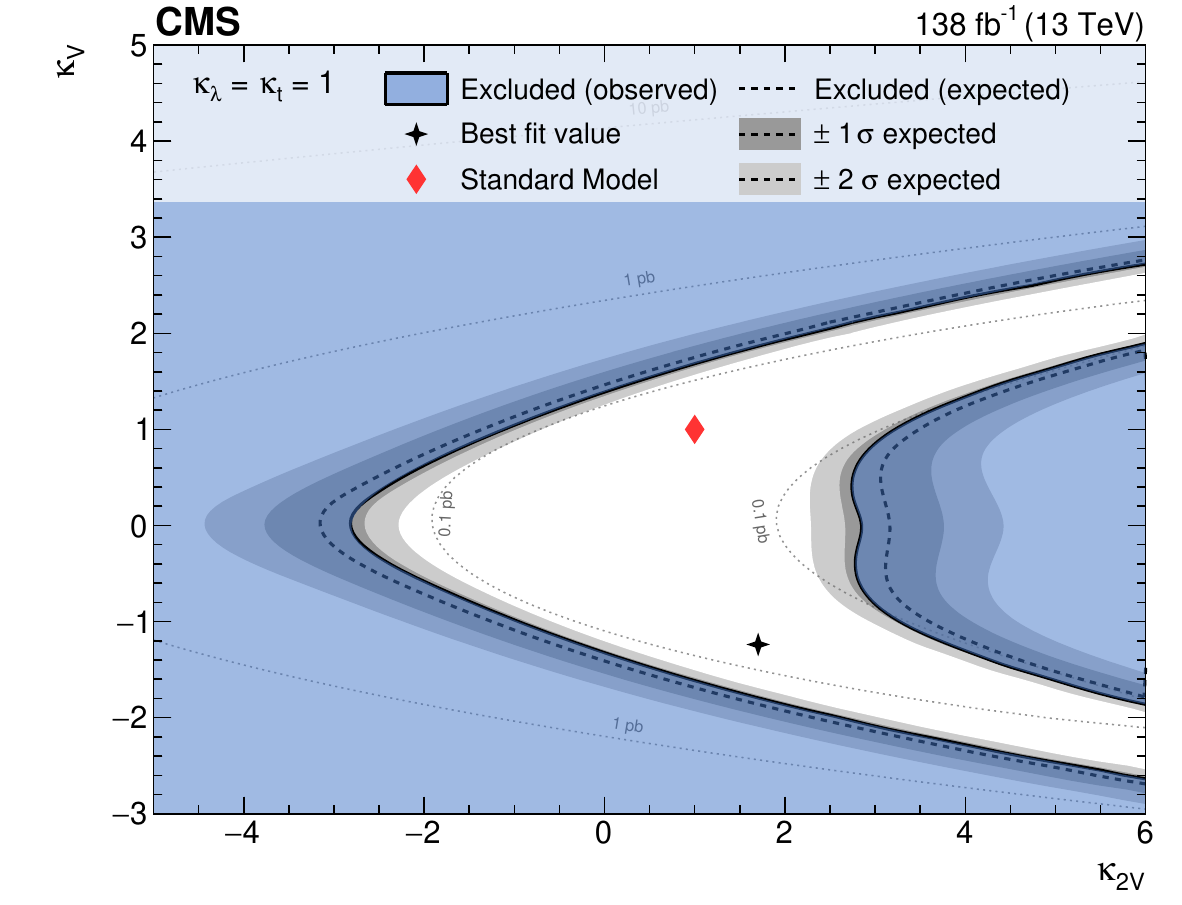}
  \caption{Observed and expected 95\% \CL exclusion limits on the nonresonant \HH production via \VBF cross section as a function of the effective couplings \CV and \CVV. The blue area is excluded by the observation. The 1 and $2\sigma$ confidence intervals around the expected median exclusion contour are shown as dark and light-grey areas corresponding to 68 and 95\% respectively. The red diamond shows the SM expectation while the fine dashed lines show the theoretical cross section contours. The \GGF contribution in this case is set to the SM expectation. All other Higgs boson couplings are set to the values predicted by the SM.}
  \label{fig:2D_exclusion_kv_kvv}
\end{figure}

\begin{figure}[htb!]
  \centering
  \includegraphics[width=0.75\textwidth]{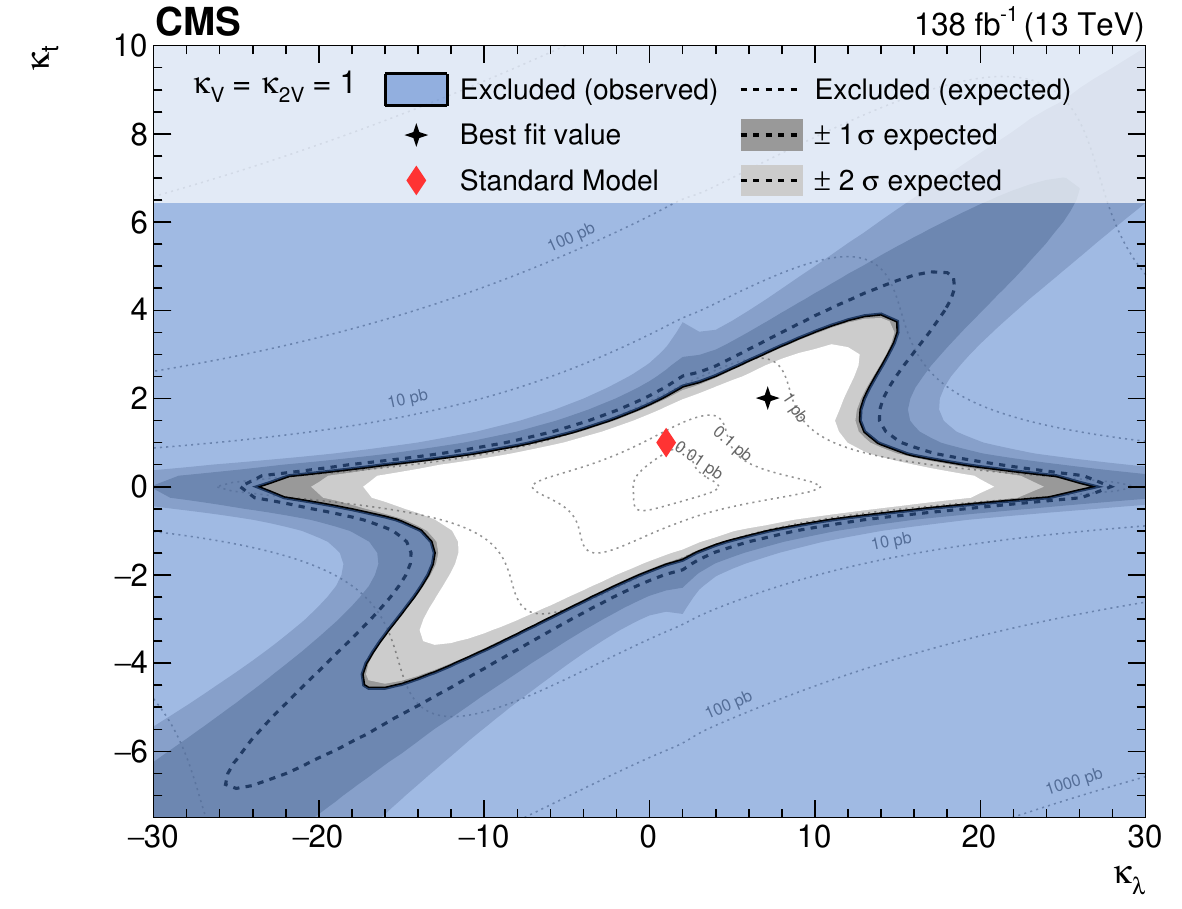}
  \caption{Observed and expected 95\% \CL exclusion limits on the nonresonant \HH production cross section as a function of the effective couplings \klambda and \ktop. The blue area is excluded by the observation. The  1 and $2\sigma$ confidence intervals around the expected median exclusion contour are shown as dark and light-grey areas corresponding to 68 and 95\% respectively. The red diamond shows the SM expectation while the fine dashed lines show the theoretical cross section contours. All other Higgs boson couplings are set to the values predicted by the SM.}
  \label{fig:2D_exclusion_kt_kl}
\end{figure}

The exclusion contours on the \klambda and \CVV, \CV and \CVV, \ktop and \klambda  are shown in Figs.~\ref{fig:2D_exclusion_kl_kvv}, \ref{fig:2D_exclusion_kv_kvv}, and \ref{fig:2D_exclusion_kt_kl}, respectively, while all other Higgs boson couplings are set to the values predicted by the SM.

\begin{figure}[htb!]
  \centering
  \includegraphics[width=0.75\textwidth]{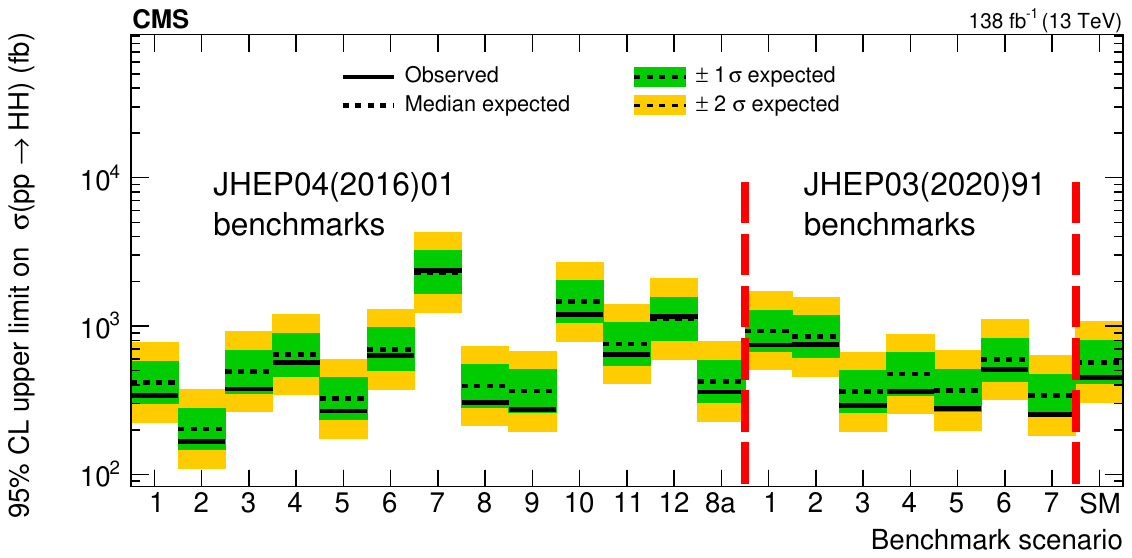}
  \caption{Observed and expected 95\% \CL upper limits on the nonresonant \HH production cross section for two different benchmark scenarios ``JHEP04(2016)01'' and ``JHEP03(2020)91'' from Refs.~\cite{Carvalho_2016,Capozi_2020}. The green and yellow bands show the 1 and $2\sigma$ confidence intervals, corresponding to 68 and 95\% \CL.}
  \label{fig:benchmarks_all}
\end{figure}

\begin{figure}[htb!]
  \centering
  \includegraphics[width=0.75\textwidth]{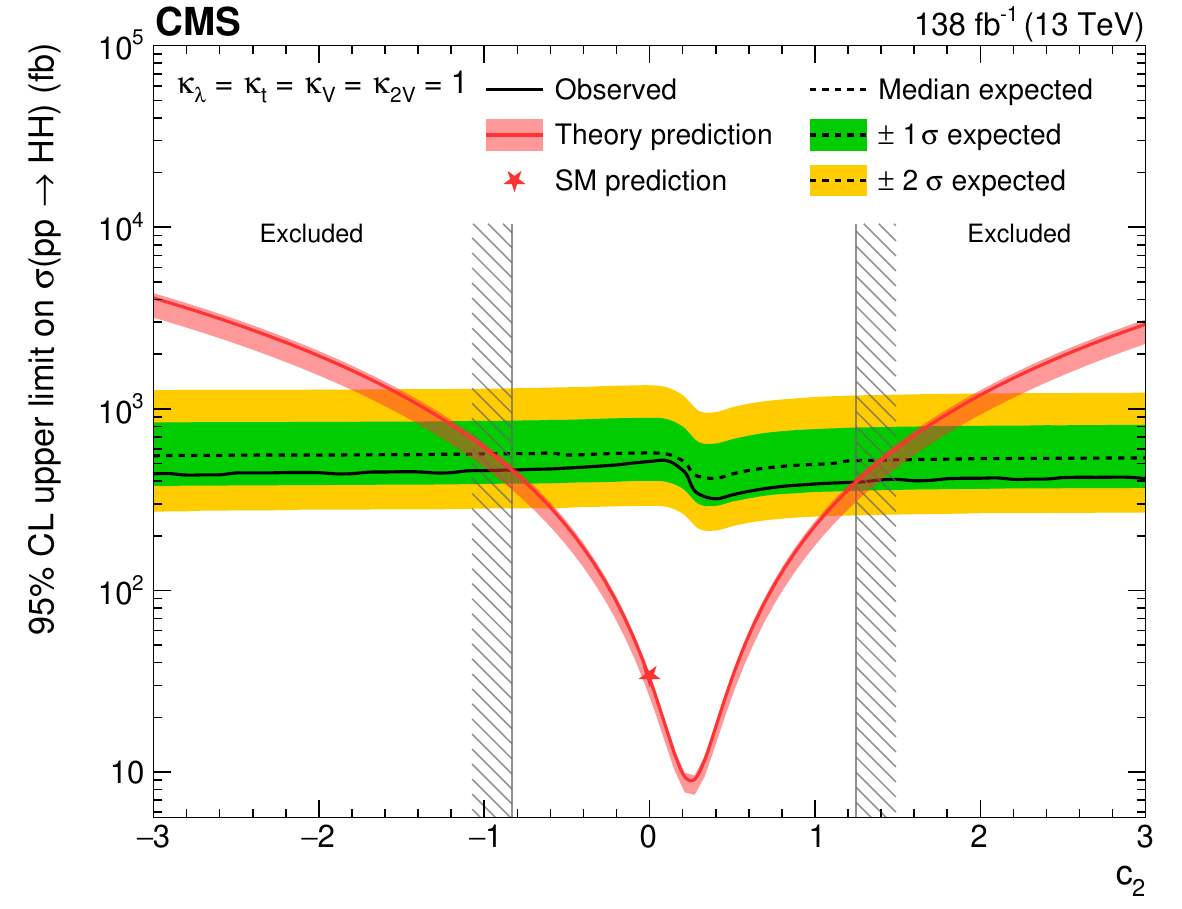}
  \caption{Observed and expected 95\% CL upper limits on the nonresonant \HH production cross section as a function of the effective coupling \ctwo. The green and yellow bands show the 1 and $2\sigma$ confidence intervals, corresponding to 68 and 95\% \CL. All other Higgs boson couplings are set to the values predicted in the SM. Overlaid in red (upper) is the curve representing the predicted \HH production cross section.}
  \label{fig:c2_limits_1D}
\end{figure}

\begin{figure}[htb!]
  \centering
  \includegraphics[width=0.75\textwidth]{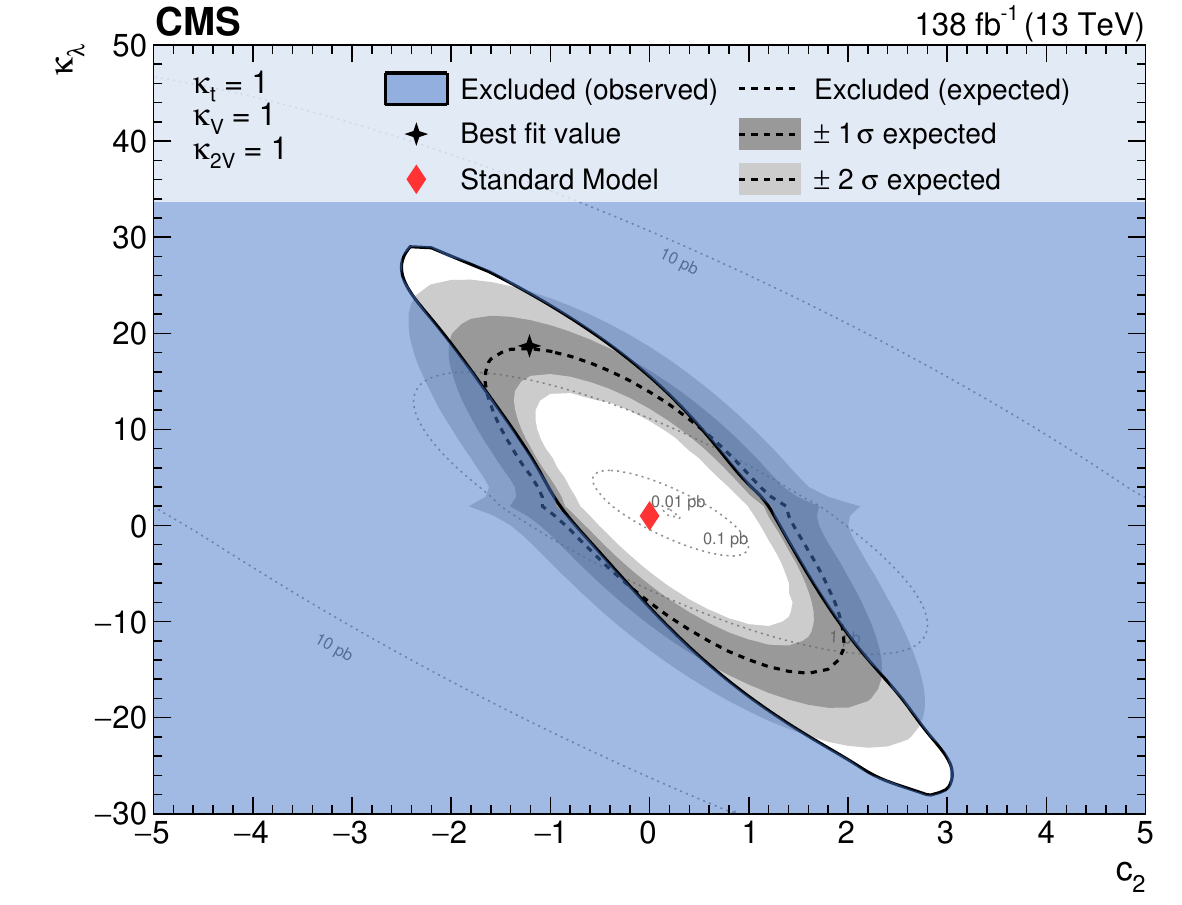}
  \caption{Observed and expected 95\% CL exclusion limits on the nonresonant \HH production cross section as a function of the effective couplings \klambda and \ctwo. The blue area is excluded by the observation. The 1 and $2\sigma$ confidence intervals around the expected median exclusion contour are shown as dark and light-grey areas corresponding to 68 and 95\% respectively. The red diamond shows the SM expectation while the fine dashed lines show the theoretical cross section contours. All other Higgs boson couplings are set to the values predicted in the SM.}
  \label{fig:c2_limits_2D}
\end{figure}

As explained in Section~\ref{sec:HHsimulation}, we can study modified \HH couplings as well as anomalous couplings which are not predicted in the SM. The set of coupling modifiers studied is \klambda, \ktop, \ctwo, \cg and \cgg (Figs.~\ref{fig:dihiggs-production-diagrams-ggf_SM} and \ref{fig:dihiggs-production-diagrams-ggf_BSM}), based on an effective field theory parameterisation. Two sets of benchmarks referred to as ``JHEP04(2016)01''~\cite{Carvalho_2016} and ``JHEP03(2020)91''~\cite{Capozi_2020} have been proposed,
which are different combinations of (\klambda, \ktop, \ctwo, \cg, \cgg). The results are interpreted as limits on the \GGF production cross section for each of the benchmarks considered, shown in Fig.~\ref{fig:benchmarks_all}. In addition, a limit scan for the \ctwo coupling is shown in Fig.~\ref{fig:c2_limits_1D}. The \ctwo coupling is constrained between $[-0.8,\ 1.3]$ (expected $[-1.0,1.4]$) at 95\% \CL. Exclusion limits contours are drawn in the \klambda-\ctwo plane, shown in Fig.~\ref{fig:c2_limits_2D}. All other Higgs boson couplings are set to the values predicted by the SM.

\begin{figure}[htb!]
  \centering
  \includegraphics[width=0.75\textwidth]{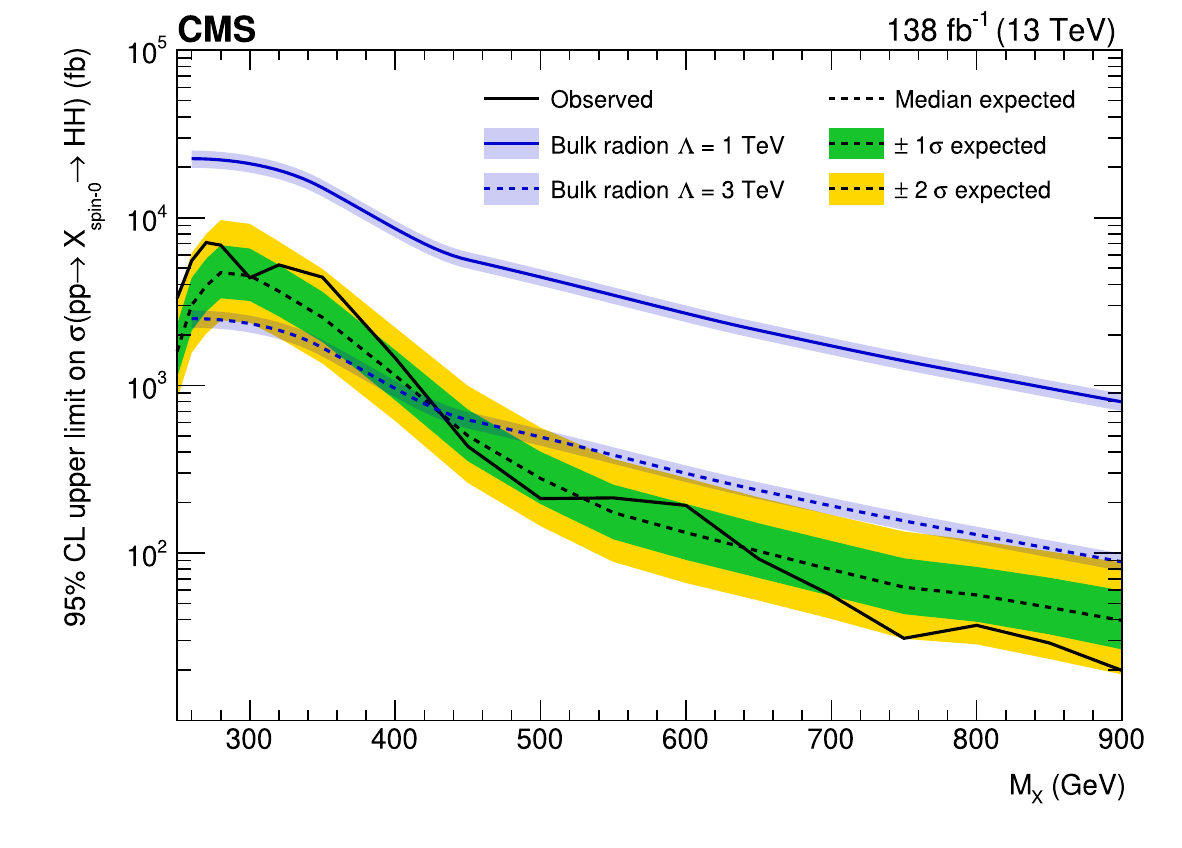}\\
  \includegraphics[width=0.75\textwidth]{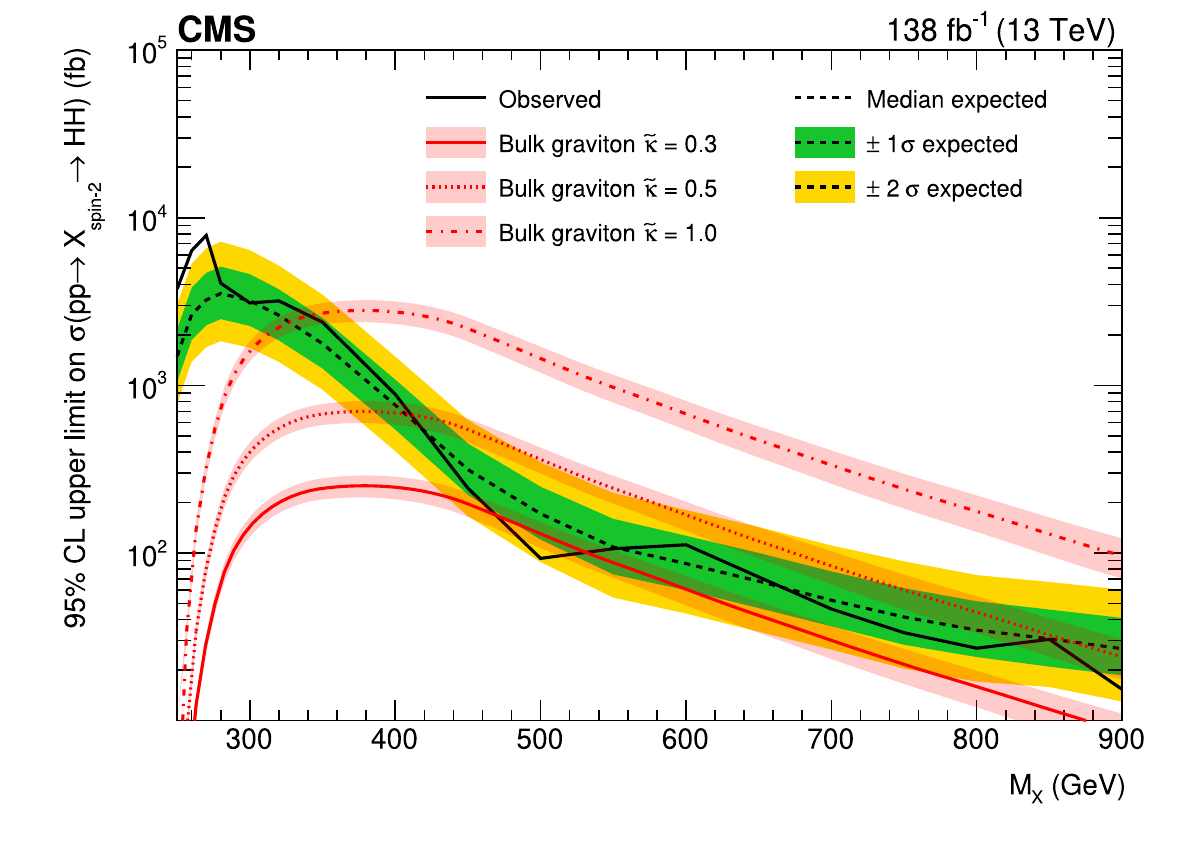}
  \caption{Observed and expected 95\% CL upper limits on the production of new particles $X$ of spin-0 (upper) and spin-2 (lower) and mass $m_X$ in the range $250 \leq m_X \leq 900\GeV$, which
    decay to Higgs boson pairs. The green and yellow bands show the 1 and $2\sigma$ confidence intervals, corresponding to 68 and 95\% \CL. Theory predictions in benchmark scenarios for bulk radion (upper) and bulk graviton (lower) models are overlaid.}
  \label{fig:resonant_limits}
\end{figure}

The results of the resonant search are extracted using the same strategy and similar categories to the nonresonant one.
The limits are presented as a function of the heavy resonance mass hypothesis in the range between 250 and 900\GeV, since 250\GeV is the threshold for on-shell \HH production. Beyond 900\GeV the topology of the Higgs boson decays becomes predominantly highly boosted and the analysis presented in this paper is not optimal. A previously published CMS search covers this phase-space~\cite{CMSbbWWboosted2022}. The resonances are assumed to have a narrow width, \ie width smaller than the experimental resolution for the reconstructed Higgs boson mass which is 10--15\%. Figure~\ref{fig:resonant_limits} shows the limits for the spin-0 and spin-2 signal hypotheses. For the spin-0 (2) scenario the limits on the cross section vary between 5540 (6368) and 20 (15)\fb corresponding to the 250\GeV and 900\GeV mass points, respectively. For resonant mass above 700\GeV the limits set by the \bbWW search are comparable to those by the \bbgg search~\cite{CMS_HHbbgg_R2Resonant}.
Theoretical predictions for the spin-0 radion and the spin-2 graviton are shown against the respective exclusion limits. The parameters $\Lambda$ and $\tilde{\kappa} = \kappa/M_{\text{Pl}}$ correspond to the energy scale and warp factor of the so-called ``bulk" benchmark of the warped extra dimensions scenarios~\cite{Carvalho:2014lsg,Gouzevitch:1536639}, where $M_{\text{Pl}}$ is the Planck mass. The values of these parameters have been chosen based on the current constraints from other measurements.
While these are representative models for this type of search, the search is performed in a model-independent way, only depending on the resonance mass, width, and spin.

\clearpage

\section{Summary}
\label{summary}
In this paper, a search for Higgs boson pair production (\HH) in the $\HH\to\bbWW$ decay channel  with at least one electron or muon in the final state is presented.
The nonresonant and the resonant production mechanisms are studied.
No significant deviation from the standard model (SM) background is found. Upper limits are set on the \HH production cross section.

The cross section for the inclusive nonresonant $\HH\to\bbWW$ production is excluded up to a minimum of 14 (expected 18) times the value predicted by the SM at 95\% confidence level.
Compared to previous results on the same process by the CMS Collaboration,
this search represents a significant improvement with a gain in sensitivity by up to a factor of five.
The vector boson fusion production is excluded up to 277 (expected 301) times the value predicted by the SM at 95\% confidence level.

The limits on the cross sections are also shown as a function of various Higgs boson coupling modifiers and anomalous Higgs boson couplings.
The Higgs boson trilinear coupling \lambdahhh is constrained between $[-7.2,13.8]$ (expected $[-8.7,15.2]$) times the value expected in the SM at 95\% confidence level. The coupling modifier for the quartic interaction between two Higgs bosons and two \PW or \PZ bosons, \CVV, is constrained between $[-1.1,3.2]$ (expected $[-1.4,3.5]$) at 95\% confidence level. The coupling between two top quarks and two Higgs bosons, which is predicted to be zero in the SM, is constrained between $[-0.8,\ 1.3]$ (expected $[-1.0,1.4]$) at 95\% confidence level. The exclusion contours are drawn as a function of the Higgs boson coupling modifiers.

The \HH production via a heavy resonance is studied in the mass range from 250 to 900\GeV.
Spin-0 and spin-2 scenarios for the resonance are tested and compared to the common theoretical benchmarks of a heavy $CP$-even scalar radion and a graviton. The limits on the resonance production cross section in the spin-0 (2) scenario vary between 5540 (6368) and 20 (15)\fb corresponding to the 250 and 900\GeV mass points, respectively. These limits are comparable to those set by the \bbgg resonant \HH search.  

\begin{acknowledgments}
We congratulate our colleagues in the CERN accelerator departments for the excellent performance of the LHC and thank the technical and administrative staffs at CERN and at other CMS institutes for their contributions to the success of the CMS effort. In addition, we gratefully acknowledge the computing centres and personnel of the Worldwide LHC Computing Grid and other centres for delivering so effectively the computing infrastructure essential to our analyses. Finally, we acknowledge the enduring support for the construction and operation of the LHC, the CMS detector, and the supporting computing infrastructure provided by the following funding agencies: SC (Armenia), BMBWF and FWF (Austria); FNRS and FWO (Belgium); CNPq, CAPES, FAPERJ, FAPERGS, and FAPESP (Brazil); MES and BNSF (Bulgaria); CERN; CAS, MoST, and NSFC (China); MINCIENCIAS (Colombia); MSES and CSF (Croatia); RIF (Cyprus); SENESCYT (Ecuador); ERC PRG, RVTT3 and MoER TK202 (Estonia); Academy of Finland, MEC, and HIP (Finland); CEA and CNRS/IN2P3 (France); SRNSF (Georgia); BMBF, DFG, and HGF (Germany); GSRI (Greece); NKFIH (Hungary); DAE and DST (India); IPM (Iran); SFI (Ireland); INFN (Italy); MSIP and NRF (Republic of Korea); MES (Latvia); LMTLT (Lithuania); MOE and UM (Malaysia); BUAP, CINVESTAV, CONACYT, LNS, SEP, and UASLP-FAI (Mexico); MOS (Montenegro); MBIE (New Zealand); PAEC (Pakistan); MES and NSC (Poland); FCT (Portugal); MESTD (Serbia); MCIN/AEI and PCTI (Spain); MOSTR (Sri Lanka); Swiss Funding Agencies (Switzerland); MST (Taipei); MHESI and NSTDA (Thailand); TUBITAK and TENMAK (Turkey); NASU (Ukraine); STFC (United Kingdom); DOE and NSF (USA).

\hyphenation{Rachada-pisek} Individuals have received support from the Marie-Curie programme and the European Research Council and Horizon 2020 Grant, contract Nos.\ 675440, 724704, 752730, 758316, 765710, 824093, 101115353,101002207, and COST Action CA16108 (European Union); the Leventis Foundation; the Alfred P.\ Sloan Foundation; the Alexander von Humboldt Foundation; the Science Committee, project no. 22rl-037 (Armenia); the Belgian Federal Science Policy Office; the Fonds pour la Formation \`a la Recherche dans l'Industrie et dans l'Agriculture (FRIA-Belgium); the Agentschap voor Innovatie door Wetenschap en Technologie (IWT-Belgium); the F.R.S.-FNRS and FWO (Belgium) under the ``Excellence of Science -- EOS" -- be.h project n.\ 30820817; the Beijing Municipal Science \& Technology Commission, No. Z191100007219010 and Fundamental Research Funds for the Central Universities (China); the Ministry of Education, Youth and Sports (MEYS) of the Czech Republic; the Shota Rustaveli National Science Foundation, grant FR-22-985 (Georgia); the Deutsche Forschungsgemeinschaft (DFG), under Germany's Excellence Strategy -- EXC 2121 ``Quantum Universe" -- 390833306, and under project number 400140256 - GRK2497; the Hellenic Foundation for Research and Innovation (HFRI), Project Number 2288 (Greece); the Hungarian Academy of Sciences, the New National Excellence Program - \'UNKP, the NKFIH research grants K 124845, K 124850, K 128713, K 128786, K 129058, K 131991, K 133046, K 138136, K 143460, K 143477, 2020-2.2.1-ED-2021-00181, and TKP2021-NKTA-64 (Hungary); the Council of Science and Industrial Research, India; ICSC -- National Research Centre for High Performance Computing, Big Data and Quantum Computing, funded by the EU NexGeneration program (Italy); the Latvian Council of Science; the Ministry of Education and Science, project no. 2022/WK/14, and the National Science Center, contracts Opus 2021/41/B/ST2/01369 and 2021/43/B/ST2/01552 (Poland); the Funda\c{c}\~ao para a Ci\^encia e a Tecnologia, grant CEECIND/01334/2018 (Portugal); the National Priorities Research Program by Qatar National Research Fund; MCIN/AEI/10.13039/501100011033, ERDF ``a way of making Europe", and the Programa Estatal de Fomento de la Investigaci{\'o}n Cient{\'i}fica y T{\'e}cnica de Excelencia Mar\'{\i}a de Maeztu, grant MDM-2017-0765 and Programa Severo Ochoa del Principado de Asturias (Spain); the Chulalongkorn Academic into Its 2nd Century Project Advancement Project, and the National Science, Research and Innovation Fund via the Program Management Unit for Human Resources \& Institutional Development, Research and Innovation, grant B37G660013 (Thailand); the Kavli Foundation; the Nvidia Corporation; the SuperMicro Corporation; the Welch Foundation, contract C-1845; and the Weston Havens Foundation (USA).
\end{acknowledgments}

\bibliography{auto_generated}

\cleardoublepage \appendix\section{The CMS Collaboration \label{app:collab}}\begin{sloppypar}\hyphenpenalty=5000\widowpenalty=500\clubpenalty=5000
\cmsinstitute{Yerevan Physics Institute, Yerevan, Armenia}
{\tolerance=6000
A.~Hayrapetyan, A.~Tumasyan\cmsAuthorMark{1}\cmsorcid{0009-0000-0684-6742}
\par}
\cmsinstitute{Institut f\"{u}r Hochenergiephysik, Vienna, Austria}
{\tolerance=6000
W.~Adam\cmsorcid{0000-0001-9099-4341}, J.W.~Andrejkovic, T.~Bergauer\cmsorcid{0000-0002-5786-0293}, S.~Chatterjee\cmsorcid{0000-0003-2660-0349}, K.~Damanakis\cmsorcid{0000-0001-5389-2872}, M.~Dragicevic\cmsorcid{0000-0003-1967-6783}, A.~Escalante~Del~Valle\cmsorcid{0000-0002-9702-6359}, P.S.~Hussain\cmsorcid{0000-0002-4825-5278}, M.~Jeitler\cmsAuthorMark{2}\cmsorcid{0000-0002-5141-9560}, N.~Krammer\cmsorcid{0000-0002-0548-0985}, D.~Liko\cmsorcid{0000-0002-3380-473X}, I.~Mikulec\cmsorcid{0000-0003-0385-2746}, J.~Schieck\cmsAuthorMark{2}\cmsorcid{0000-0002-1058-8093}, R.~Sch\"{o}fbeck\cmsorcid{0000-0002-2332-8784}, D.~Schwarz\cmsorcid{0000-0002-3821-7331}, M.~Sonawane\cmsorcid{0000-0003-0510-7010}, S.~Templ\cmsorcid{0000-0003-3137-5692}, W.~Waltenberger\cmsorcid{0000-0002-6215-7228}, C.-E.~Wulz\cmsAuthorMark{2}\cmsorcid{0000-0001-9226-5812}
% ...
\par}
\vskip\cmsinstskip
\dag:~Deceased\\
$^{1}$Also at Yerevan State University, Yerevan, Armenia\\
$^{2}$Also at TU Wien, Vienna, Austria\\
% ...
\end{sloppypar}
\end{document}